\documentclass[journal=jctcce,manuscript=article,layout=traditional]{achemso}
\setkeys{acs}{doi = true, etalmode = firstonly, maxauthors = 15, abbreviations = true}
\usepackage{graphicx}% Include figure files
\usepackage{dcolumn}% Align table columns on decimal point
\usepackage{bm}% bold math
\usepackage{verbatim}
\usepackage{soul}
\usepackage{float}
\usepackage{xcolor}
\usepackage{chemformula} % Formula subscripts using \ch{}
\usepackage[T1]{fontenc} % Use modern font encodings
\usepackage{multirow}
\usepackage[colorlinks = true, allcolors = blue,]{hyperref}% add hypertext capabilities
\usepackage{orcidlink}

\author{Diem Thi-Xuan Dang\,\orcidlink{0000-0001-7136-4125}}
\affiliation{Department of Physics, University of South Florida, Tampa, Florida 33620, USA}
\author{Yen Thi-Hai Pham}
\affiliation{Department of Physics, University of South Florida, Tampa, Florida 33620, USA}
\altaffiliation{Current address: Department of Physics and Astronomy, George Mason University, Fairfax, Virginia 22030, USA}
\author{Da Zhou}
\affiliation{Department of Physics, The Pennsylvania State University, University Park, Pennsylvania 16802, USA}
\alsoaffiliation{Center for 2- Dimensional and Layered Materials, The Pennsylvania State University, University Park, Pennsylvania 16802, USA}
\author{Dai-Nam Le\,\orcidlink{0000-0003-0756-8742}}
\affiliation{Department of Physics, University of South Florida, Tampa, Florida 33620, USA}
\author{Mauricio Terrones}
\affiliation{Department of Physics, The Pennsylvania State University, University Park, Pennsylvania 16802, USA}
\alsoaffiliation{Center for 2- Dimensional and Layered Materials, The Pennsylvania State University, University Park, Pennsylvania 16802, USA}
\author{Manh-Huong Phan\,\orcidlink{0000-0002-6270-8990}}
\email{phanm@usf.edu}
\affiliation{Department of Physics, University of South Florida, Tampa, Florida 33620, USA}
\author{Lilia M. Woods\,\orcidlink{0000-0002-9872-1847}}
\email{lmwoods@usf.edu}
\affiliation{Department of Physics, University of South Florida, Tampa, Florida 33620, USA}

\title{Interface Magnetism in Vanadium-doped \ch{MoS2}/Graphene Heterostructures}

\begin{document}

\begin{abstract}
Magnetism in two-dimensional materials is of great importance in discovering new physical phenomena and developing new devices at the nanoscale. In this paper, first-principles simulations are used to calculate the electronic and magnetic properties of heterostructures composed of Graphene and \ch{MoS2} considering the influence of point defects and Vanadium doping. It is found that the concentration of the dopants and the types of defects can result in induced magnetic moments leading to ferromagnetically polarized systems with sharp interfaces. This provides a framework for interpreting the experimental observations of enhanced ferromagnetism in both \ch{MoS2}/Graphene and V-doped \ch{MoS2}/Graphene heterostructures.
The computed electronic and spin polarizations give a microscopic understanding of the origin of ferromagnetism in these systems and illustrate how doping and defect engineering can lead to targeted property tunability. Our work has demonstrated that through defects engineering, ferromagnetism can be achieved in V-doped \ch{MoS2}/Graphene heterostructures, providing a potential way to induce magnetization in other TMDC/Graphene materials and opening new opportunities for their applications in nano-spintronics.

\end{abstract}
\maketitle

\section{\label{sec:1}Introduction}

The properties of low-dimensional materials and heterostructures (HSTs), in which relatively weak dispersion interactions play an important role, are strongly modified by the proximity effect. A prominent example is Graphene, which experiences relatively strong proximity spin–orbit coupling (SOC) effects when placed on transition metal dichalcogenides (TMDC). Low-dimensional HSTs are currently used in optospintronics \cite{Luo2017, Avsar2017}, as well as in a variety of applications where the spin Hall effect and the spin galvanic effect are prominent \cite{Ghiasi2019, Benitez2020}.

Typically, TMDC/Graphene HSTs are non-magnetic. Nevertheless, the development of methods to effectively generate and control magnetism in 2D materials is critical for the establishment of new magnetic HSTs with various applications. An effective method to induce magnetic properties in 2D semiconductors involves doping and defect engineering \cite{Zhang2020, Coelho2024}. Based on previous studies, transition metal substitutional doping can effectively generate tunable magnetism in TMDC systems \cite{Pham2020, Ortiz2021}. Among them, Vanadium doping can realize not only p-type transport behavior in TMDCs but also room-temperature ferromagnetism, making V-doped TMDCs a favorable platform for electronic and spintronic devices \cite{Yun2020,Nguyen2023}. 

Recently, V-doped \ch{MoS2} has been intensively studied. Zhang et al. demonstrated that V-doped \ch{MoS2} increases radiative recombination of excitons relative to trions by reducing background electrons in \ch{MoS2} \cite{Zhang2020_2}. Sahoo et al. reported the presence of time-reversal asymmetry between the two K-valleys with a 32 meV relative energy shift between them \cite{Sahoo2022}. Maity et al. observed a semiconductor-to-metal transition in V-doped \ch{MoS2} alloy when the doping concentration increased above 5$\%$ \cite{Maity2022}. The magnetic performance of the V-doped \ch{MoS2} can be tuned by varying the doping concentration. In particular, enhanced ferromagnetism is observed for up to 3$\%$ V doping concentration \cite{Zhou2024}, which is strongly inhibited by raising doping beyond 8$\%$ \cite{Zhou2024}. However, due to the low doping concentration, the net magnetization of these monolayers remains relatively weak limiting their suitability for spintronic and magnetic sensor applications.

Constructing a HST of a doped TMDC exhibiting ferromagnetic behavior and a nonmagnetic 2D system such as Graphene brings forward new possibilities for magnetic HSTs with sharp interfaces. Of special interest is to see if Graphene can become magnetic as a result of interactions and proximity effects with the ferromagnetic TMDC \cite{Tan2018,Cai2021}. In this study, we broaden our understanding of doped TMDCs and their magnetism by investigating V-doped \ch{MoS2}/Graphene HSTs by taking into account various types of defects and doping concentrations. In Graphene, mono- and divacancies, Stone-Wales defects, etc. have been studied quite extensively \cite{Nakhmedov2019, Tiwari2023}. The fabrication of TMDCs can also result in various types of vacancies \cite{Lin2016, Han2024}, including V-doped \ch{MoS2} monolayer obtained by chemical vapor deposition methods \cite{Zhou2024}. These vacancies are known to have dramatic effects on the electronic, magnetic, and optical properties of TMDC systems \cite{Khan2017, Koos2019, Zhou2024}. Due to their importance, in this paper, we have systematically studied the role of point defects in the properties of the V-doped \ch{MoS2}/Graphene HST. We present various types of defect formations, the electronic structures of the defective materials, and the defect signatures of the magnetic properties of the different HSTs. Our work shows pathways to improve the saturation magnetization of TMDC/Graphene HSTs, which can be useful for their application in spintronics.
 
\section{\label{sec:2} Results and Discussions}
\subsection{Structural properties}

The primary investigation system is an HST formed by stacking a single Graphene layer (Gr) and a single \ch{MoS2} layer. To initiate the calculations, we construct an atomistic supercell consisting of 162 C atoms ($9 \times 9 \times 1$ Gr unit cell) and 49 Mo, 98 S atoms ($7 \times 7 \times 1$ \ch{MoS2} unit cell). After relaxation, the in-plane lattice constant of the Gr/\ch{MoS2} HST is $a = b = 22.183$ \AA. As a reference, each monolayer is also simulated separately with an optimized lattice constant for Gr found as $a = b = 2.471$ \AA, while the optimized lattice constant for \ch{MoS2} is $a = b = 3.153$ \AA. 

In addition to a HST composed of pristine monolayers, several point-like defects in both monolayers are considered, because isolated point defects are unavoidable in the synthesis process as discussed earlier. Specifically, a C monovacancy (Gr+\ch{V_C}), C divacancies (Gr+\ch{V_{2C}}) and Stone-Wales (Gr+\ch{V_{SW}}) defect are constructed in the Gr part of the supercell. The defective TMDC monolayers are also generated in the \ch{MoS2} part of the supercell and they are: \ch{MoS2}+\ch{V_S} (with a S monovacancy), \ch{MoS2}+\ch{V_{S-S}} (with two adjacent S vacancies on the same S layer), \ch{MoS2}+\ch{V_{2S}} (with two S vacancies on the same Mo atom but in opposite S layers), \ch{MoS2}+\ch{V_{Mo}} (with a Mo monovacancy) and \ch{MoS2}+\ch{V_{Mo+2S}} (with a Mo vacancy and two S vacancies on opposite layers). To probe the magnetism in the system, Vanadium substitutional dopants at the Mo sites in the TMDC layer are simulated by taking \ch{V_{0.02}Mo_{0.98}S2} ($2\%$ V doping) and \ch{V_{0.04}Mo_{0.96}S2} ($4\%$ V doping) monolayers. Various combinations for the combined effect of point defects and dopants within the \ch{MoS2} layer are also taken by constructing \ch{V_{0.02}Mo_{0.98}S2}+\ch{V_S} ($2\%$ V doping and a S monovacancy), \ch{V_{0.02}Mo_{0.98}S2}+\ch{V_{S-S}} ($2\%$ V doping and two adjacent S vacancies on the same layer), \ch{V_{0.02}Mo_{0.98}S2}+\ch{V_{2S}} ($2\%$ V doping and two adjacent S vacancies on the different layer), \ch{V_{0.02}Mo_{0.98}S2}+\ch{V_{Mo}} ($2\%$ V doping and a Mo monovacancy), and \ch{V_{0.02}Mo_{0.98}S2}+\ch{V_{Mo+2S}} (with $2\%$ V doping, a Mo vacancy and two adjacent S vacancies on the different layers) monolayers. 

\begin{figure}[H]
    \begin{center}
    \includegraphics[width = 1 \textwidth]{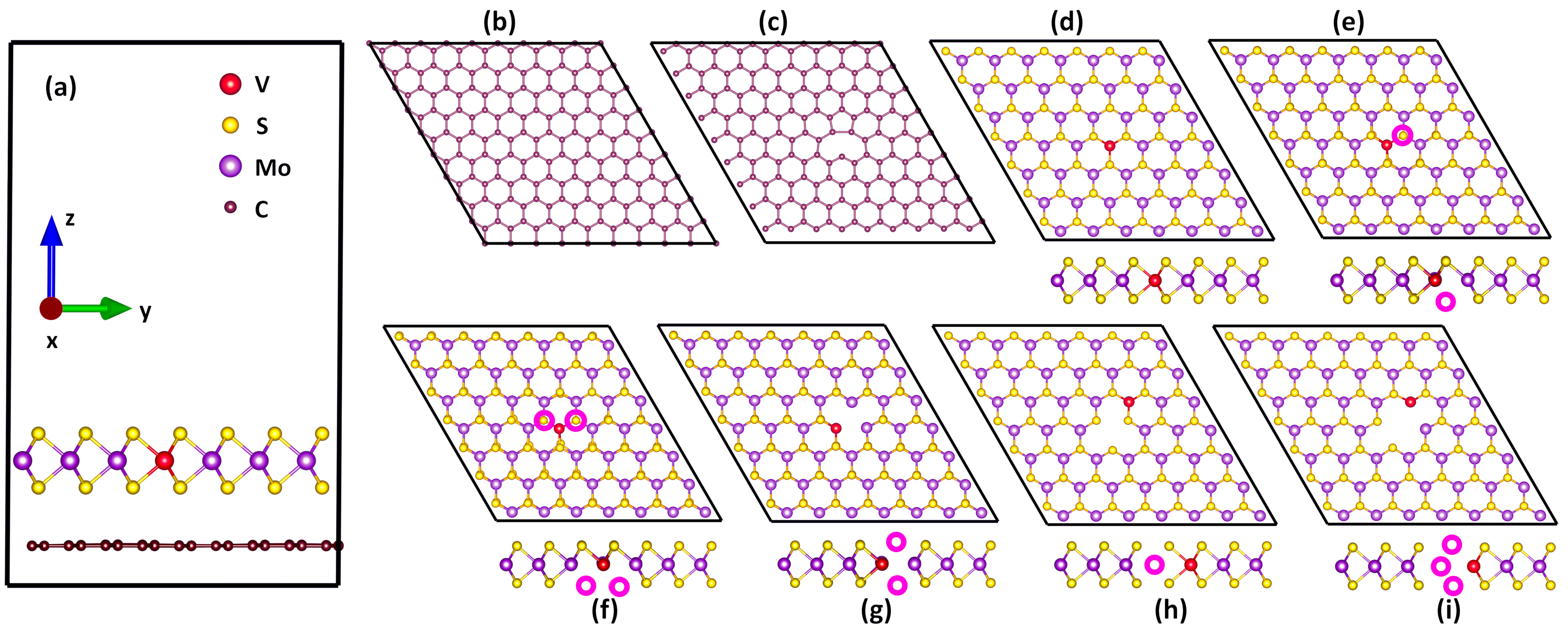}
    \caption{\label{fig:1} (a) Side view of the pristine HST showing labels for the different atoms and the interlayer separation $d$. Top views of the Gr monolayers as part of the V-doped \ch{MoS2}/Gr HSTs: (b) pristine Gr, (c) Gr+\ch{V_C} (with C monovacancy). Top and side views of the various \ch{V_{0.02}Mo_{0.98}S2} monolayers as part of the V-doped \ch{MoS2}/Gr HSTs: (d) \ch{V_{0.02}Mo_{0.98}S2} (with $2\%$ V substitutional doping), (e) \ch{V_{0.02}Mo_{0.98}S2}+\ch{V_S} (with $2\%$ V substitutional doping and S monovacancy), (f) \ch{V_{0.02}Mo_{0.98}S2}+\ch{V_{S-S}} (with $2\%$ V doping and S divacancies on the same atomic layer), (g) \ch{V_{0.02}Mo_{0.98}S2}+\ch{V_{2S}} (with $2\%$ V substitutional doping and S divacancies on different atomic layers), (h) \ch{V_{0.02}Mo_{0.98}S2}+\ch{V_{Mo}} (with $2\%$ substitutional V doping and Mo monovacancy), (i) \ch{V_{0.02}Mo_{0.98}S2}+\ch{V_{Mo+2S}} (with $2\%$ V doping and an Mo+2S defect complex).}
    \end{center}
\end{figure}

In this study, we simulate HSTs composed of various combinations of Gr and TMDC sheets discussed above. We find that the magnetic properties of HSTs composed of different TMDCs and pristine Gr or Gr+\ch{V_{2C}} or Gr+\ch{V_{SW}} are very similar due to the similar nonmagnetic nature of these Gr sheets. Thus, in what follows, we only consider HSTs composed of different TMDCs and pristine Gr or Gr+\ch{V_C}. The full list of the simulated HSTs composed of various combinations of defective and/or doped monolayers is given in Table \ref{tab:1}, and several representative examples of the lattice configurations are shown in Fig. \ref{fig:1}. The numerous combinations (26 structures altogether) help us to understand the complex interplay between the types of point defects in a specific monolayer and the magnetic dopants in the Gr/\ch{MoS2} HSTs.

Table \ref{tab:1} shows the lattice mismatch of each HST by giving the strain each layer experiences in the supercell after full relaxation. It appears that upon the formation of the different HSTs, the Gr layer is shrunk while the TMDC layer is expanded, but in all cases the effect is less than a percent. For example, \ch{V_{0.02}Mo_{0.98}S2} sheet in Gr/\ch{V_{0.02}Mo_{0.98}S2} HST is stretched by $0.426\%$, whereas the Gr sheet is compressed by $0.247\%$ from the optimized pristine cell parameters. \ch{V_{0.02}Mo_{0.98}S2} sheet in (Gr+\ch{V_C)/V_{0.02}Mo_{0.98}S2} HST is stretched by $0.462\%$, while the Gr sheet is compressed by $0.234\%$ from the optimized parameters of the pristine cell. These rather small values indicate that the built-in strain will be a negligible factor in the overall behavior of the simulated HSTs. 

\begin{table}[H]
\caption{\label{tab:1} Built-in strain in each monolayer calculated using $\epsilon = \dfrac{a_{HST} - a_i}{a_i}  \times 100 \% \; $  ($a_{HST}$ - lattice constant of the HST; $a_i$ - lattice constant of the Gr or TMDC layers)}
\resizebox{\textwidth}{!}{
\begin{tabular}{|l|c|c|l|c|c|}
\hline\hline
Name & Gr Strain $(\%)$ & \ch{MoS2} Strain $(\%)$ & Name & Gr Strain  $(\%)$ & \ch{MoS2} Strain $(\%)$ \\ \hline
HST1: Gr/\ch{MoS2} & -0.234 & 0.498 & HST14: (Gr+\ch{V_C})/\ch{MoS2} & -0.225 & 0.530 \\ \hline
HST2: Gr/(\ch{MoS2}+\ch{V_S}) & -0.283 & 0.644 & HST15: (Gr+\ch{V_C})/(\ch{MoS2}+\ch{V_S}) & -0.288 & 0.654 \\ \hline
HST3: Gr/(\ch{MoS2}+\ch{V_{S-S}}) & -0.337 & 0.824 & HST16: (Gr+\ch{V_C})/(\ch{MoS2}+\ch{V_{S-S}}) & -0.351 & 0.833 \\ \hline
HST4: Gr/(\ch{MoS2}+\ch{V_{2S}}) & -0.342 & 0.778 & HST17: (Gr+\ch{V_C})/(\ch{MoS2}+\ch{V_{2S}}) & -0.342 & 0.782 \\ \hline
HST5: Gr/(\ch{MoS2}+\ch{V_{Mo}}) & -0.220 & 0.489 & HST18: (Gr+\ch{V_C})/(\ch{MoS2}+\ch{V_{Mo}}) & -0.197 & 0.534 \\ \hline
HST6: Gr/(\ch{MoS2}+\ch{V_{Mo+2S}}) & -0.382 & 0.806 & HST19: (Gr+\ch{V_C})/(\ch{MoS2}+\ch{V_{Mo+2S}}) & -0.697 & 0.708 \\ \hline
HST7: Gr/\ch{V_{0.02}Mo_{0.98}S2} & -0.247 & 0.426 & HST20: (Gr+\ch{V_C})/\ch{V_{0.02}Mo_{0.98}S2} & -0.234 & 0.462 \\ \hline
HST8: Gr/\ch{V_{0.04}Mo_{0.96}S2} & -0.229 & 0.389 & HST21: (Gr+\ch{V_C})/\ch{V_{0.04}Mo_{0.96}S2} & -0.499 & 0.407 \\ \hline
HST9: Gr/(\ch{V_{0.02}Mo_{0.98}S2}+\ch{V_S}) & -0.288 & 0.535 & HST22: (Gr+\ch{V_C})/(\ch{V_{0.02}Mo_{0.98}S2}+\ch{V_S}) & -0.306 & 0.512 \\ \hline
HST10: Gr/(\ch{V_{0.02}Mo_{0.98}S2}+\ch{V_{S-S}}) & -0.351 & 0.723 & HST23: (Gr+\ch{V_C})/(\ch{V_{0.02}Mo_{0.98}S2}+\ch{V_{S-S}}) & -0.373 & 0.723 \\ \hline
HST11: Gr/(\ch{V_{0.02}Mo_{0.98}S2}+\ch{V_{2S}}) & -0.346 & 0.714 & HST24: (Gr+\ch{V_C})/(\ch{V_{0.02}Mo_{0.98}S2}+\ch{V_{2S}}) & -0.364 & 0.695 \\ \hline
HST12: Gr/(\ch{V_{0.02}Mo_{0.98}S2}+\ch{V_{Mo}}) & -0.234 & 0.631 & HST25: (Gr+\ch{V_C})/(\ch{V_{0.02}Mo_{0.98}S2}+\ch{V_{Mo}}) & -0.184 & 0.480 \\ \hline
HST13: Gr/(\ch{V_{0.02}Mo_{0.98}S2}+\ch{V_{Mo+2S}}) &  -0.292 & 0.800 & HST26: (Gr+\ch{V_C})/(\ch{V_{0.02}Mo_{0.98}S2}+\ch{V_{Mo+2S}}) & -0.306 & 0.809 \\ \hline\hline
\end{tabular}%
}
\end{table}

\begin{figure}[H]
    \begin{center}
    \includegraphics[width = 0.55 \textwidth]{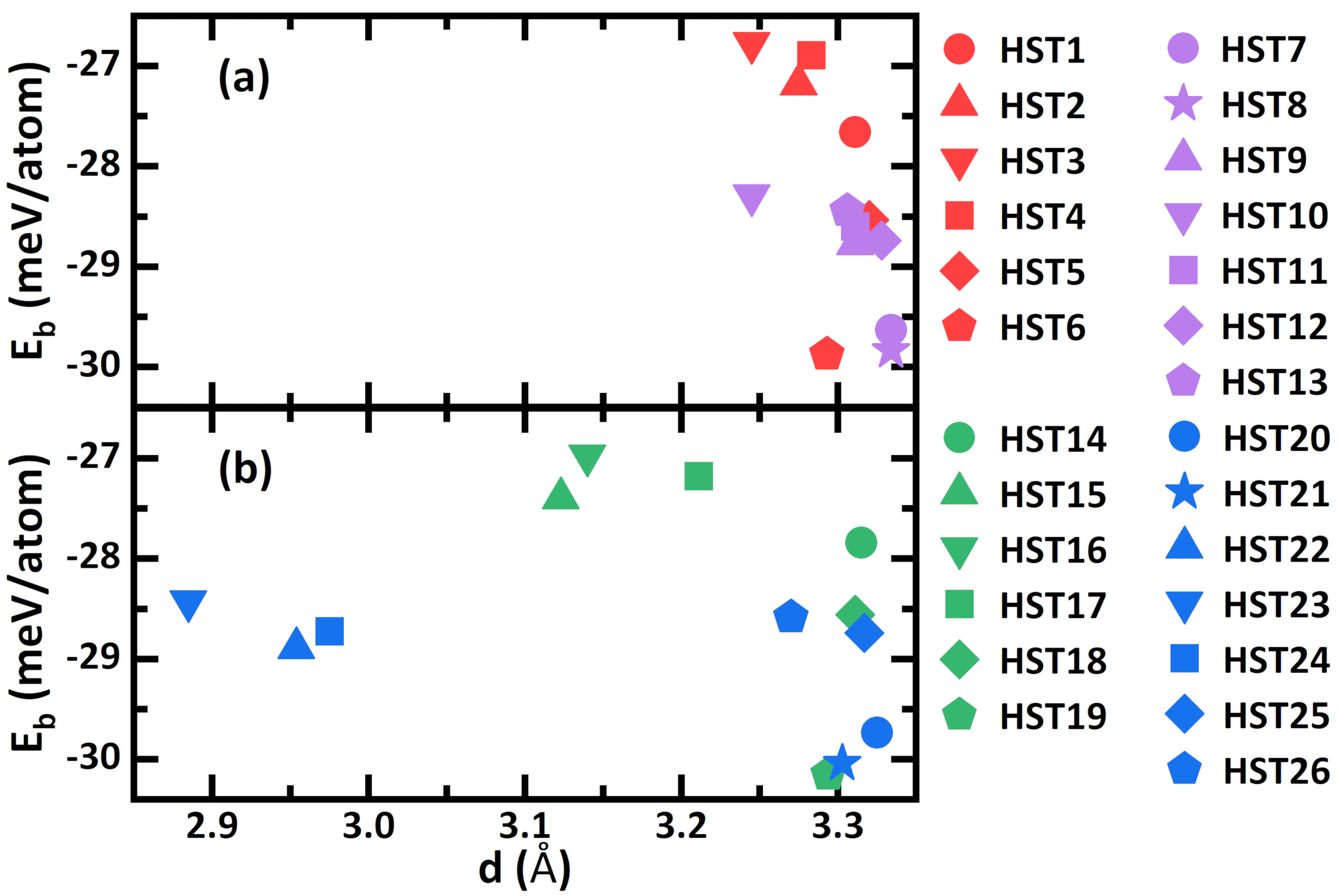}
    \caption{\label{fig:2} Interlayer binding energy $E_b$ (eV) as a function of the interlayer separation $d$ (\AA) \text{} of HSTs: (a) Gr/TMDCs and (b) (Gr+\ch{V_C})/TMDCs.}
    \end{center}
\end{figure}

To gain some insight into the interface properties of the HSTs, we compute the interlayer binding energy using the relation $E_b = (E_{HST} - E_{L1} - E_{L2})/N$, where $E_{HST}$ is the total energy of each HST, $E_{L1, L2}$ are the total energies for the individual monolayers, and $N$ is the number of atoms in the system. Fig. \ref{fig:2} displays $E_b$ versus the interlayer separation $d$ for all systems. The results show that the binding energy is in the range $E_b = (-30.152, -26.772)$ meV/atom, while the interlayer separation varies in $d = (2.885, 3.334)$ \AA \text{} range. The stability of all HSTs is reflected in the negative values of $E_b$, and we find that those systems containing \ch{MoS2} with a S monovacancy or divacancies have the highest binding energy. On the other hand, the lowest binding energy is found for systems containing \ch{MoS2} with a Mo$+$2S vacancy complex or V dopants, indicating that such HSTs have a better structural stability compared to the rest of the systems. These $E_b$ values are typical for HSTs in which van der Waals (vdW) interactions play an important role in their structural stability \cite{Dang2025}.  

The interlayer separation in all cases is also in the typical vdW range, since for all systems $d$ is found to be greater than the sum of the covalent radii of C and S atoms (0.76 and 1.05 \AA, respectively). For Gr/\ch{V_{0.04}Mo_{0.96}S2} HST, $d$ is very similar to the case of the pristine Gr/\ch{V_{0.02}Mo_{0.98}S2} HST. However, it appears that $d$ between Gr and defective \ch{V_{0.02}Mo_{0.98}S2} monolayers is smaller as compared to the one for the pristine Gr/\ch{V_{0.02}Mo_{0.98}S2} HST. This effect becomes more pronounced for the defective Gr. 

\subsection{Magnetic properties}

\begin{figure}[H]
    \begin{center}
    \includegraphics[width = 0.45 \textwidth]{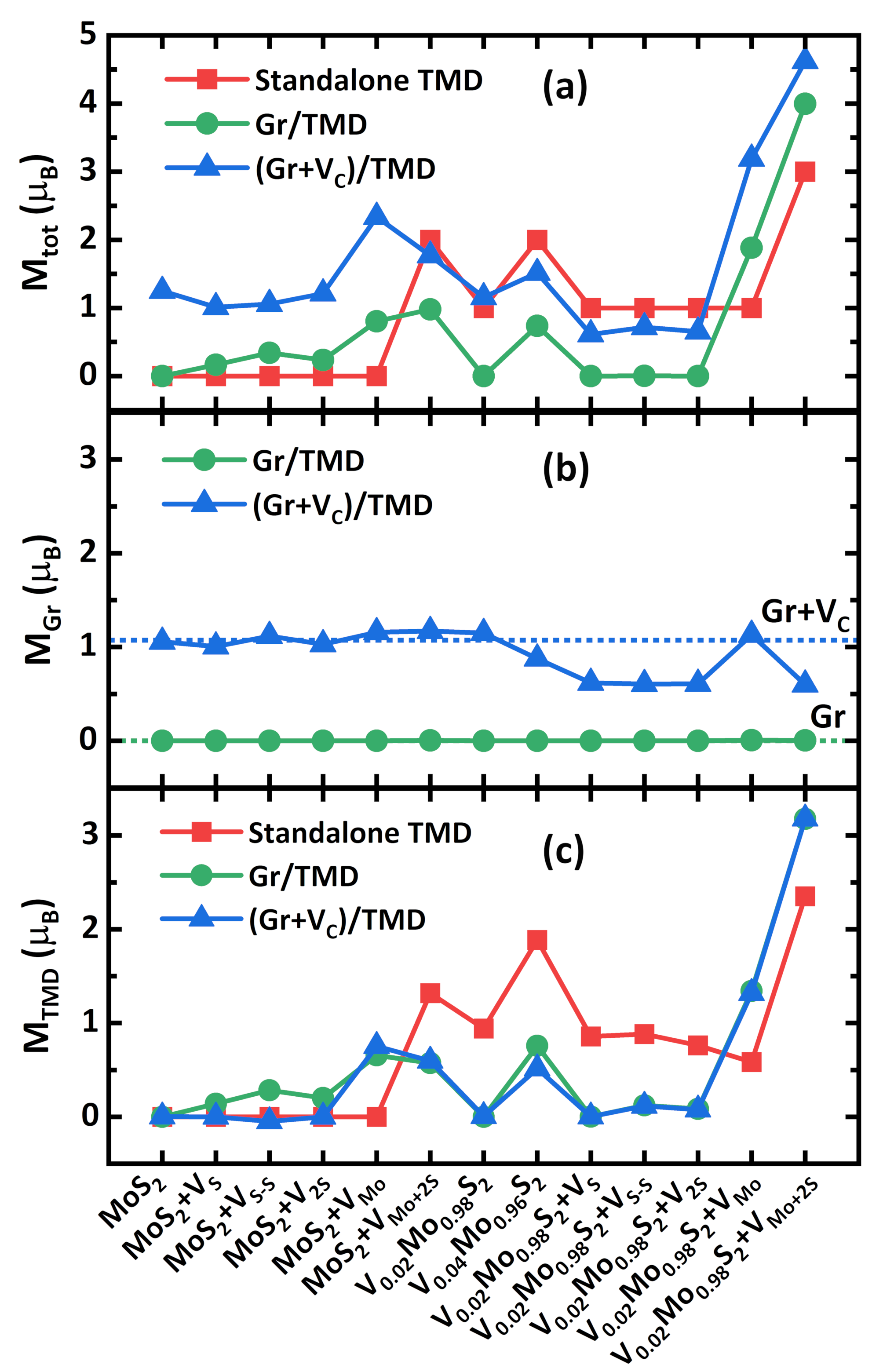}
    \caption{\label{fig:3} Total magnetization of the: (a) entire HST $M_{tot}$ $(\mu_B)$, (b) Gr layer $M_{Gr}$ $(\mu_B)$, (c) TMDC layer $M_{TMDC}$ $(\mu_B)$ as part of HSTs. The data is given as a function of the different TMDC used to construct each HST. }
    \end{center}
\end{figure}

The interplay between defects and dopants also affects the magnetic properties of the studied HSTs. Our calculations show that several materials exhibit a ferromagnetic ordering, and in Fig. \ref{fig:3}  we show results for the total magnetization of the system $M_{tot}$, and the total magnetization for the corresponding Gr layer, $M_{Gr}$, and TMDC layer, $M_{TMDC}$. 

Let us first examine the isolated \ch{MoS2} monolayer. From the calculations, we find that isolated S monovacancy and divacancies, and Mo monovacancy do not induce magnetic ordering  in the isolated TMDC monolayer. This agrees with previous studies on defective and single-doped \ch{MoS2} \cite{Wang2016}. The total magnetic moment is 2 $\mu_B$ for \ch{MoS2}+\ch{V_{Mo+2S}}. On the other hand, $M_{tot} = 1$ $\mu_B$ for a V-doped \ch{MoS2} with or without S monovacancy, S divacancies, Mo monovacancy (\ch{V_{0.02}Mo_{0.98}S2}, \ch{V_{0.02}Mo_{0.98}S2}+\ch{V_S}, \ch{V_{0.02}Mo_{0.98}S2}+\ch{V_{S-S}}, \ch{V_{0.02}Mo_{0.98}S2}+\ch{V_{2S}}, \ch{V_{0.02}Mo_{0.98}S2}+\ch{V_{Mo}}). Increasing of the doping concentration raises $M_{tot}$ to 2 $\mu_B$ for \ch{V_{0.04}Mo_{0.96}S2}, while $M_{tot} = 3$ $\mu_B$ for \ch{V_{0.02}Mo_{0.98}S2}+\ch{V_{Mo+2S}}. Such synergistic effects have been observed experimentally \cite{Hu2019}.

This situation changes upon the HST formation by adding a Gr layer. In fact, Fig. \ref{fig:3}  shows that there is a tendency of reducing the total magnetization when a pristine Gr is stacked above a \ch{MoS2} with mono or divacancies despite their V dopants. It is interesting to observe that $M_{tot} = 1$ $\mu_B$ of the stand-alone \ch{V_{0.02}Mo_{0.98}S2} becomes $M_{tot} = 0$ for Gr/\ch{V_{0.02}Mo_{0.98}S2}. In other words, a ferromagnetic standalone \ch{MoS2} with a $2\%$ V doping becomes nonmagnetic upon bringing Gr together (HST7). Fig. \ref{fig:3} also shows that this trend holds for \ch{Gr/(V_{0.02}Mo_{0.98}S2}+\ch{V_S}), \ch{Gr/(V_{0.02}Mo_{0.98}S2}+\ch{V_{S-S}}), \ch{Gr/(V_{0.02}Mo_{0.98}S2}+\ch{V_{2S})} (HST9, 10, 11), thus S mono and divacancies in the $2\%$ doped TMDC do not affect the overall lack of magnetization in the HSTs. This reduction in the total magnetization is directly associated with charge transfer from Gr to \ch{V_{0.02}Mo_{0.98}S2} layer regardless of point defect presence, as discussed below. Nevertheless, there are situations where the Gr layer enhances the FM properties of the standalone TMDC. Such is the case of a V-doped \ch{MoS2} with a Mo monovacancy or a Mo+2S complex defect. The Gr monolayer enhances the $M_{tot}$ by about $88\%$ and $33\%$ when compared to the isolated \ch{V_{0.02}Mo_{0.98}S2}+\ch{V_{Mo}} or \ch{V_{0.02}Mo_{0.98}S2}+\ch{V_{Mo+2S}}. Figs. \ref{fig:3}b, c show that in all HSTs containing a pristine Gr monolayer, the magnetic properties are carried out by the TMDC monolayer, and Gr remains nonmagnetic.

The magnetic properties of the HSTs with a defective Gr layer are overall greatly enhanced compared with those of the standalone TMDC layer and the HST with the pristine Gr layer. We find that the total magnetic moment of Gr with C monovacancy is 1.271 $\mu_B$. The FM state of (Gr+\ch{V_C}) agrees with the well-known Lieb theorem that C monovacancy induces a magnetic moment on Gr while other defects do not \cite{Lieb1989, Yazyev2007}. We further find that (Gr+\ch{V_C})/TMDC with the S point defects shows a prominent ferromagnetism with $M_{tot} > 1$ $\mu_B$ unlike the nonmagnetic TMDC monolayer or the weakly ferromagnetic Gr/TMDC system. This enhancement effect is even stronger for the (Gr+\ch{V_C})/(\ch{MoS2}+\ch{V_{Mo}}) and (Gr+\ch{V_C})/(\ch{MoS2}+\ch{V_{Mo+2S}}). It is also interesting to note that the defective Gr has elevated $M_{tot}$ for its HSTs formed with a doped TMDC with (Gr+\ch{V_C})/(\ch{V_{0.02}Mo_{0.98}S2}+\ch{V_{Mo}}) and (Gr+\ch{V_C})/(\ch{V_{0.02}Mo_{0.98}S2}+\ch{V_{Mo+2S}}) defects, for which Gr/TMDC showed weak ferromagnetism. It appears that in such cases, the magnetization is carried out primarily by the defective Gr layer (Fig. \ref{fig:3}b, c). Introducing V dopants, however, enhances the magnetization of the TMDC layer itself, while Gr+\ch{V_C} remains ferromagnetic as well. Thus, these results show that the C vacancy plays a key role in the overall magnetic properties of the entire HST because it enhances the magnetization of the Gr layer itself. Such enhancement due to the ferromagnetism of the defective Gr layer is consistent with the enhanced magnetism found in other HSTs involving TMDC monolayers and ferromagnetic layers \cite{Tan2018,Thi2022, Hung2023}.

\begin{figure}[H]
    \begin{center}
    \includegraphics[width = 1 \textwidth]{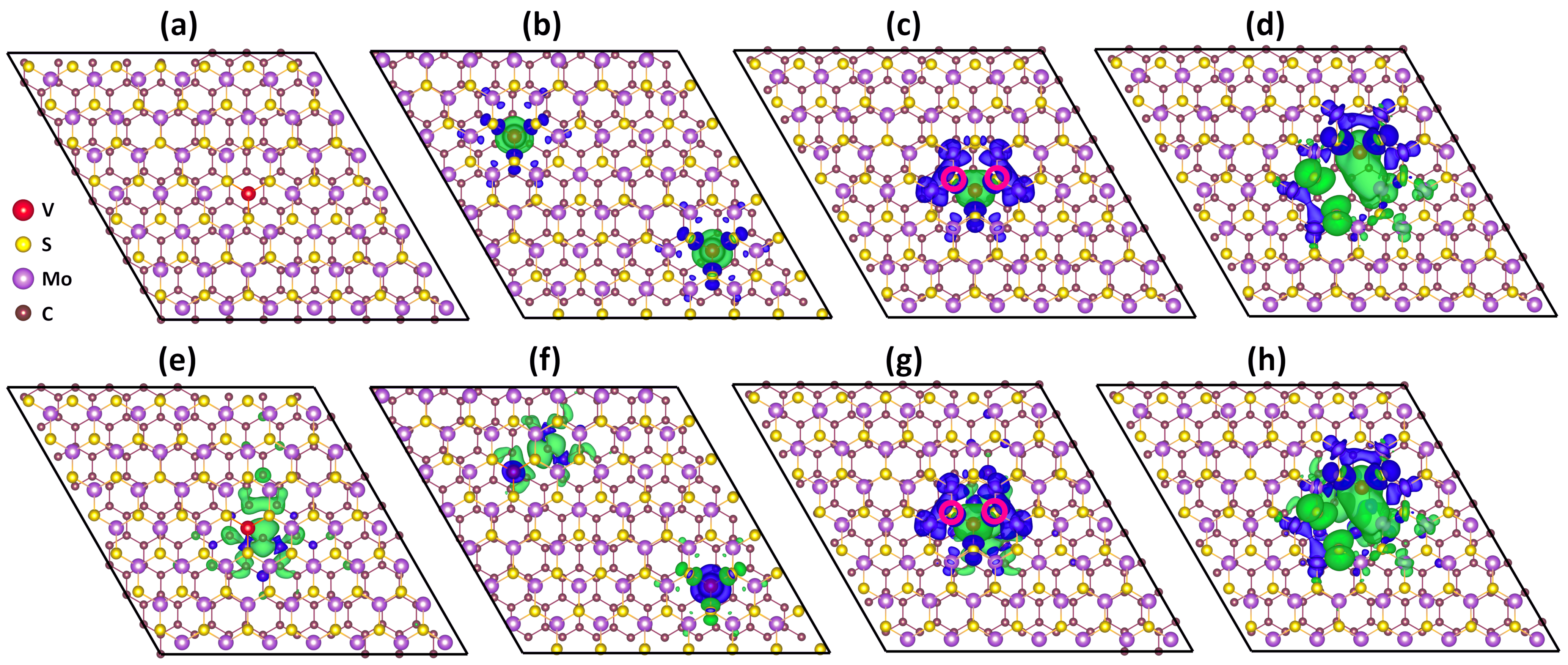}
    \caption{\label{fig:4} Electron spin densities of: (a) Gr/$\ch{V_{0.02}Mo_{0.98}S2}$, (b) Gr/\ch{V_{0.04}Mo_{0.96}S2}, (c) Gr/(\ch{V_{0.02}Mo_{0.98}S2}+\ch{V_{S-S}}), (d) Gr/(\ch{V_{0.02}Mo_{0.98}S2}+\ch{V_{Mo+2S}}), (e) (Gr+\ch{V_C})/\ch{V_{0.02}Mo_{0.98}S2}, (f) (Gr+\ch{V_C})/\ch{V_{0.04}Mo_{0.96}S2}, (g) (Gr+\ch{V_C})/(\ch{V_{0.02}Mo_{0.98}S2}+\ch{V_{S-S}}) and (h) (Gr+\ch{V_C})/(\ch{V_{0.02}Mo_{0.98}S2}+\ch{V_{Mo+2S}}) HSTs. Green and blue colors represent the spin-up and spin-down charges, respectively. The pink circles represent S vacancy positions. The isosurface value is $10^{-3}$ $e$/\AA$^3$). }
    \end{center}
\end{figure}

Further insight of the ferromagnetism of these systems can be gained by examining the electron spin density property. In Fig. \ref{fig:4}, we show some representative cases by giving the electron spin density distribution with TMDC on top of the Gr layer. \ch{Gr/V_{0.02}Mo_{0.98}S2} is non-magnetic and no spin density appears between atoms in this system (Fig. \ref{fig:4}a). The nearest Mo-site spins are antiferromagnetically coupled to the V-site spin, while the nearest S-site spins are ferromagnetically coupled to the V-site spin in \ch{V_{0.04}Mo_{0.96}S2} monolayer. However, both the nearest Mo and S-site spins couple antiferromagnetically with the V-site spin in \ch{V_{0.04}Mo_{0.96}S2} layer of Gr/\ch{V_{0.04}Mo_{0.96}S2} HST (Fig. \ref{fig:4}b). Although the Gr/(\ch{V_{0.02}Mo_{0.98}S2}+\ch{V_{S-S}}) HST has a total magnetic moment of zero, the antiferromagnetic orders appeared around the V atom $(0.807$ $\mu_B)$ in the TMDC layer (Fig. \ref{fig:4}c). In \ch{V_{0.02}Mo_{0.98}S2}+\ch{V_{Mo+2S}} sheet of \ch{Gr/(V_{0.02}Mo_{0.98}S2}+\ch{V_{Mo+2S}}) HST, the antiferromagnetic orders appeared on the nearest Mo and S atoms around V atom, while the ferromagnetic orders appeared on the nearest Mo and S atoms around vacancy. In this case, we can clearly see that the spin density is located not only on the atoms but also on the empty spaces (Fig. \ref{fig:4}d). Spin density distributions are not found in the Gr sheets of these HSTs, which are consistent with the values shown in Fig. \ref{fig:3}b. For defective Gr sheet of (Gr+\ch{V_C})/TMDC HSTs, the C monovacancy produces two pentagons while the FM order is observed in one pentagon and the AFM order is observed in another pentagon (Fig. \ref{fig:4}e-h). In (Gr+\ch{V_C)/V_{0.02}Mo_{0.98}S2} HST, no spin density is observed in \ch{V_{0.02}Mo_{0.98}S2} component (Fig. \ref{fig:4}e). In (Gr+\ch{V_C)/V_{0.04}Mo_{0.96}S2} HST, both V atoms have negative magnetic moments, but only one AFM order is observed for the nearest Mo and S atoms (Fig. \ref{fig:4}f). Interestingly, spin density distributions of \ch{V_{0.02}Mo_{0.98}S2}+\ch{V_{S-S}} or \ch{V_{0.02}Mo_{0.98}S2}+\ch{V_{Mo+2S}} are almost unchanged in HSTs with the pristine or defective Gr sheets (Figs. \ref{fig:4}g, h). 

We also examined the magnetic moments of \ch{V_{0.02}Mo_{0.98}S2} with various vacancies placed on Gr with C divacancies or SW defect and found that the magnetic phenomena in these systems are similar to those observed in pristine Gr. This is because the three systems pristine Gr, Gr+\ch{V_{2C}} and Gr+\ch{V_{SW}} are nonmagnetic.

\subsection{Electronic Structures Properties}

\begin{figure}[H]
    \begin{center}
    \includegraphics[width = 1 \textwidth]{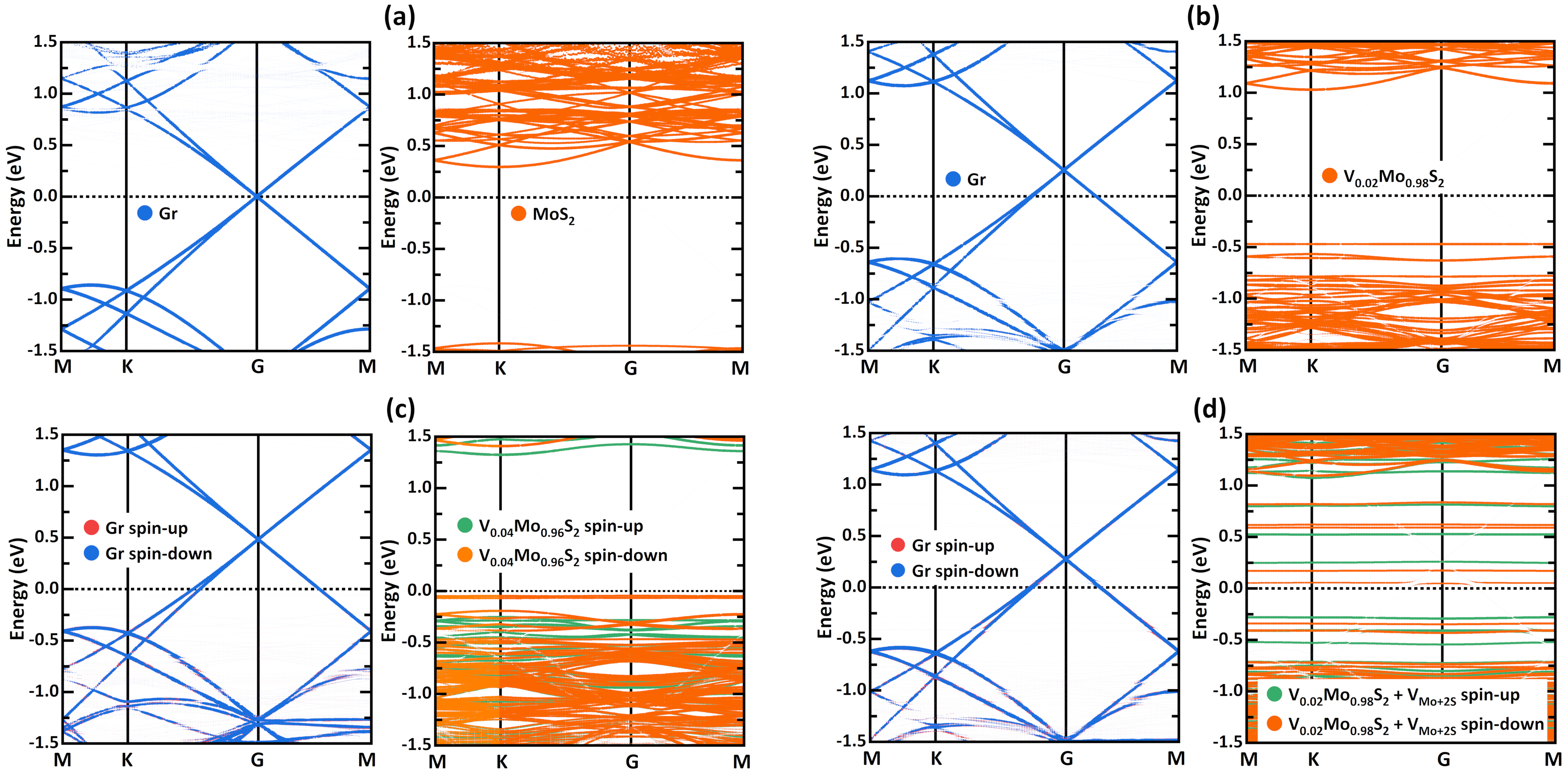}
    \caption{\label{fig:5} Band structures of (a) Gr/\ch{MoS2}, (b) Gr/\ch{V_{0.02}Mo_{0.98}S2}, (c) Gr/\ch{V_{0.04}Mo_{0.96}S2} and (d) Gr/(\ch{V_{0.02}Mo_{0.98}S2}+\ch{V_{Mo+2S}}) HSTs where the dotted lines represent the Fermi level.}
    \end{center}
\end{figure}

\begin{figure}[H]
    \begin{center}
    \includegraphics[width = 1 \textwidth]{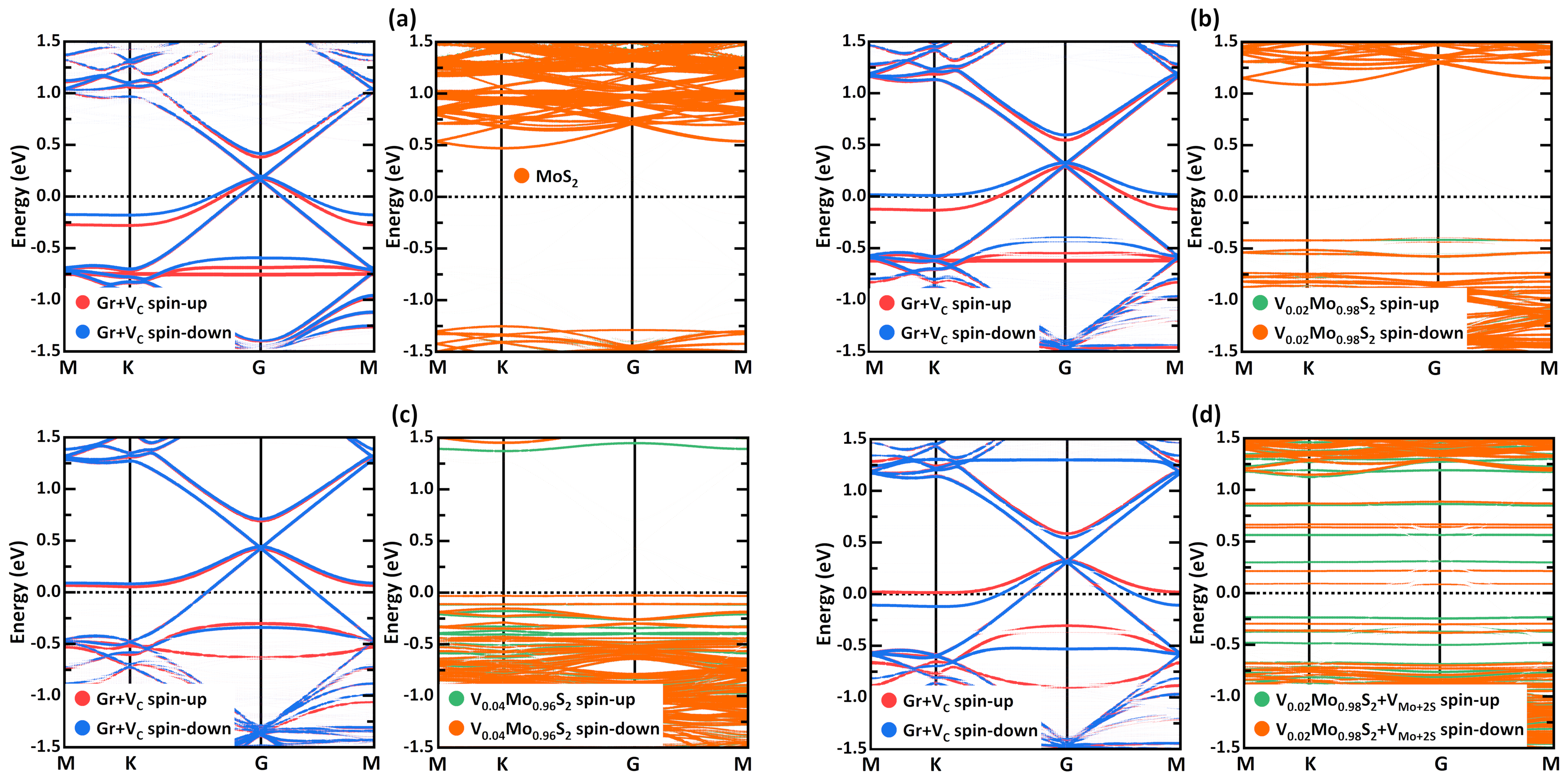}
    \caption{\label{fig:6} Band structures of (a) (Gr+\ch{V_C})/\ch{MoS2}, (b) (Gr+\ch{V_C})/\ch{V_{0.02}Mo_{0.98}S2}, (c) (Gr+\ch{V_C})/\ch{V_{0.04}Mo_{0.96}S2} and (d) (Gr+\ch{V_C})/(\ch{V_{0.02}Mo_{0.98}S2}+\ch{V_{Mo+2S}})HSTs where the dotted lines represent the Fermi level. }
    \end{center}
\end{figure}

We further discuss the electronic properties of the HSTs by analyzing their band structures (Figs. \ref{fig:5}-\ref{fig:6}). For better clarity, the band structures for the individual layers in each HST are presented.  Metallic behavior is observed for all cases except for the pristine Gr/\ch{MoS2} showing to be a semi-metal (Fig. \ref{fig:5}a). In all cases, the well-known  Dirac cones of an isolated Gr monolayer are maintained upon HST formation with the TMDC. 

Fig. \ref{fig:5}b shows that V doping is responsible for the electronic state of Gr/\ch{V_{0.02}Mo_{0.98}S2} HST becoming metallic. There is an upward shift of the electron states occupied in the valence band of the TMDC layer compared to that of the pristine HST, confirming the electron depletion from the Gr to the \ch{V_{0.02}Mo_{0.98}S2} sheet.   Increasing the concentration of V doping causes the electron states occupied in the valence band to continue to shift upward, resulting in an impurity state near the Fermi level (Fig. \ref{fig:5}c). For \ch{Gr/V_{0.02}Mo_{0.98}S2} with vacancies, the defective states appear mainly in the gap region of the \ch{V_{0.02}Mo_{0.98}S2} system (Fig. \ref{fig:5}d). 

The band structures of the defective Gr in (Gr+\ch{V_C})/TMDC HSTs show that the bands for both spin carriers cross the Fermi energy (near the $\Gamma$ point) (Fig. \ref{fig:6}). Similarly to Gr, the band structures of Gr+\ch{V_C} in these HSTs show upward shifts of the Dirac points. The position of the Dirac points above the Fermi level reflects the number of electrons transferred from the Gr+\ch{V_C} to the TMDC monolayer. A notable difference between the band structures of Gr+\ch{V_C} in these HSTs is the different locations of the $\sigma$ bands around the energy range of $-1$ to $-0.5$ eV. The band structures of the TMDC components in (Gr+\ch{V_C})/TMDC HSTs are almost unchanged compared to those in Gr/TMDC HSTs. 

Upon the introduction of the C monovacancy in Gr as part of each HST, one of linear $\pi$-$\pi^*$ crossing bands breaks its degeneracy turning it into a quadratically dispersing one (Fig. \ref{fig:6}). The spin-splitting gap between these bands depends on the V doping and vacancy concentrations in the TMDC layer. In particular, the introduction of V dopants at $2\%$ concentration (Fig. \ref{fig:6}b) increases the spin splitting gap while $4\%$ concentration of V dopants (Fig. \ref{fig:6}c) reduces the spin splitting gap when compared to the pristine TMDC (Fig. \ref{fig:6}a). In the (Gr+\ch{V_C})/\ch{V_{0.02}Mo_{0.98}S2} HST (Fig. \ref{fig:6}b), the Fermi level crosses the spin-up $\pi$ band but lies completely below the entire spin-down $\pi$ band, which causes a slight increase in the defective Gr magnetism. On the other hand, in the (Gr+\ch{V_C})/\ch{V_{0.04}Mo_{0.96}S2} HST (Fig. \ref{fig:6}c), the Fermi level lies completely below both spin-up and -down $\pi$ bands and leads to a small reduction in the defective Gr magnetism. Furthermore, when (Mo+2S) vacancies are present in the V dopants with $2\%$ concentration (Fig. \ref{fig:6}d), the splitting of $\pi$ and $\pi^*$ bands is similar to that in defect-free HST (Fig. \ref{fig:6}b). However, the (Mo+2S) vacancies in TMDC layer reverses the spin polarizations of the $\pi$ and $\pi^*$ bands. This reversal reduces the magnetism in the defective Gr layer in (Gr+\ch{V_C})/(\ch{V_{0.02}Mo_{0.98}S2}+\ch{V_{Mo+2S}}) HST compared to the corresponding defect-free (Gr+\ch{V_C})/\ch{V_{0.02}Mo_{0.98}S2} HST.

\begin{figure}[H]
    \begin{center}
    \includegraphics[width = 1 \textwidth]{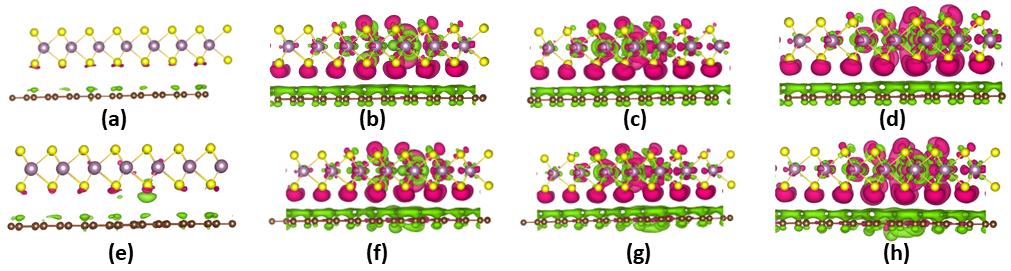}
    \caption{\label{fig:7} Charge density differences of (a) Gr/\ch{MoS2}, (b) Gr/\ch{V_{0.02}Mo_{0.98}S2}, (c) Gr/(\ch{V_{0.02}Mo_{0.98}S2}+\ch{V_{Mo+2S}}) (d) Gr/\ch{V_{0.04}Mo_{0.96}S2}, (e) (Gr+\ch{V_C})/\ch{MoS2}, (f) (Gr+\ch{V_C})/\ch{V_{0.02}Mo_{0.98}S2}, (g) (Gr+\ch{V_C})/(\ch{V_{0.02}Mo_{0.98}S2}+\ch{V_{Mo+2S}}) and (h) (Gr+\ch{V_C})/\ch{V_{0.04}Mo_{0.96}S2} HSTs. The isosurface values are $3 \times 10^{-4}$ $e$/\AA$^{-3}$. The pink isosurfaces represent the charge accumulation, the green isosurfaces represent the charge depletion.}
    \end{center}
\end{figure}

The charge transfer in each HST is closely related to its electronic structure; thus, we also calculate the 3D charge density differences in each case. The results are shown in Fig. \ref{fig:7}. We find that generally electrons are transferred from Gr or (Gr+\ch{V_C}) to the TMDC monolayer. No significant charge transfer is observed between Gr and \ch{MoS2} (Fig. \ref{fig:7}a), which is also found experimentally \cite{Pierucci2016}.  Significant charge transfer is observed between the Gr and \ch{V_{0.02}Mo_{0.98}S2}  (Fig. \ref{fig:7}b).
 The creation of vacancies or the increase in the doping concentration of V also significantly increases the charge transfer between the Gr and the TMDC layers (Figs. \ref{fig:7}c, d). This is consistent with previous theoretical and experimental papers in other HSTs involving V-doped TMDCs (\ch{Fe/W_{0.967}V_{0.033}Se2/Pt}, \ch{Bi2O2Se/WS2-V}) \cite{Thi2022, Chitara2023}. The charge transfers between the defective Gr and TMDCs are comparatively higher than those observed between the pristine Gr and TMDCs (Figs. \ref{fig:7}e-h). We can see that a significant portion of the charge density is localized on the C atoms near the defect sites, which results in a stronger interaction between the defective Gr and TMDC HSTs.

\subsection{Experimental observation of enhanced magnetization in V-doped \ch{MoS2}/Graphene heterostructures}

\begin{figure}[H]
    \begin{center}
    \includegraphics[width = 1 \textwidth]{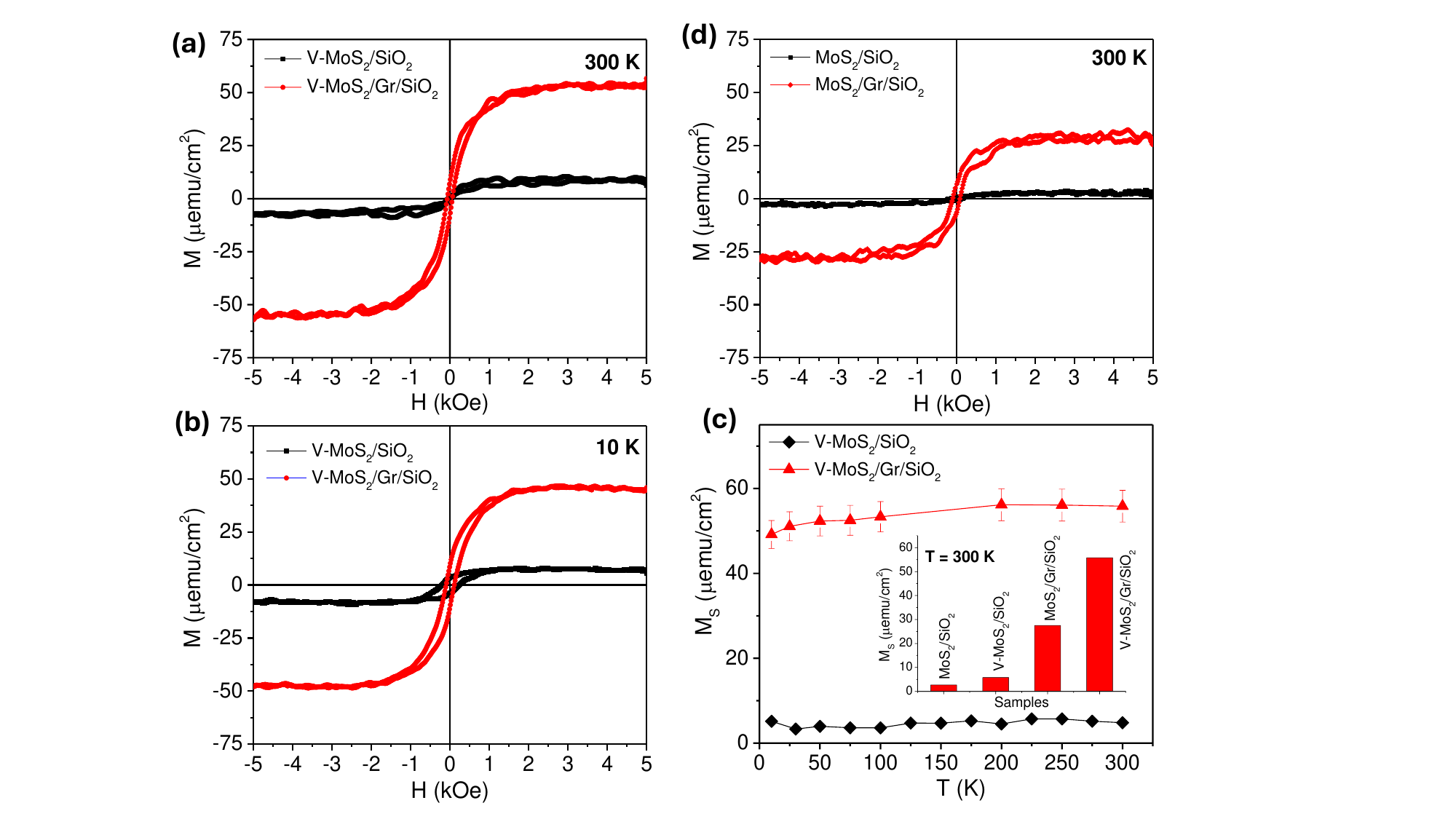}
    \caption{\label{fig:8} Magnetic hysteresis loops of the V-doped \ch{MoS2} and V-doped \ch{MoS2}/Gr samples at (a) 300 K and (b) 10 K; (c) Temperature dependence of saturation magnetization for the V-doped \ch{MoS2} and V-doped \ch{MoS2}/Gr samples; (d) Magnetic hysteresis loops taken at 300 K for the \ch{MoS2} and \ch{MoS2}/Gr samples. }
    \end{center}
\end{figure}

To experimentally validate the DFT calculations of increased magnetization in the defective and doped \ch{MoS2}/Gr HSTs, we also report measurements of the V-doped \ch{MoS2} monolayers and V-doped \ch{MoS2}/Gr HSTs alongside reference samples including pristine \ch{MoS2} monolayers and \ch{MoS2}/Gr system. Details on the growth processes of these samples are provided in the Experimental Methods section. Magnetic measurements were carried out using a vibrating sample magnetometry (VSM) probe integrated into a physical property measurement system (PPMS) over a temperature range of 10–300 K. The results are given in Fig. \ref{fig:8}(a-d). We observed that shown $2 \%$ Vanadium doping induces FM ordering at room temperature, resulting in enhanced saturation magnetization $M_S$ in the V-doped \ch{MoS2} (Fig. \ref{fig:8}(a)) compared to the pristine \ch{MoS2} monolayer (Fig. \ref{fig:8}(d)). As expected, pristine \ch{MoS2} monolayers grown on Si/SiO$_2$ substrates exhibit a weak FM signal, likely originating from various intrinsic vacancy defects \cite{Cai2015,Zhou2024}. Our DFT calculations support this, showing that specific defect configurations (e.g. \ch{V_{Mo+2S}} vacancies) dominantly induce FM ordering in the \ch{MoS2} monolayer. Other studies have also shown that the \ch{MoS2} net magnetic moment increases proportionally with the density of \ch{V_{Mo+2S}} vacancies \cite{Anbalagan2023}. Experiments have indeed demonstrated the presence of S and/or Mo vacancies in CVD-grown \ch{MoS2} and V-doped \ch{MoS2} monolayers \cite{Zhou2024,Hu2019}.

Notably, when the V-doped \ch{MoS2} monolayer is stacked with Graphene, the $M_S$ increases dramatically in the HST compared to the V-doped monolayer alone (Fig. \ref{fig:8}(a,b)). At room temperature $M_S$ increases from $\sim 5.67$ $\mu$emu/cm$^2$ in the doped TMDC to $\sim 55.83$ $\mu$emu/cm$^2$ in the HST. As shown in Fig. \ref{fig:8}(c), this enhancement persists across the full temperature range of 10 - 300 K. Moreover, the $M_S$ of the V-doped  \ch{MoS2}/Gr system  (Fig. \ref{fig:8}(a)) is nearly doubled of the undoped \ch{MoS2}/Gr (Fig. \ref{fig:8}(d)). This significant increase in the magnetization upon interfacing \ch{MoS2} with Graphene is also consistent with reports on  TMDC/Gr systems \cite{Cai2021,Tan2018}. For example, Cai et al \cite{Cai2021} suggest that covalent interactions at the \ch{MoS2} Moiré superlattice/graphene oxide interface, such as the formation of Mo–S–C bonds, lead to spin polarization of Mo 4d electrons near the Fermi level, thereby inducing high-temperature FM ordering in the HST. However, such interfacial effects alone cannot explain the substantial $M_S$ enhancement observed in our \ch{MoS2}/Gr systems (Fig. \ref{fig:8}(a,d)). The study  \cite{Tan2018} has demonstrated that vacancy defects play a critical role in inducing ferromagnetic ordering in the \ch{MoS2} monolayer as part of Mn-doped \ch{MoS2}/Gr systems. 

A comparison between Fig. \ref{fig:8}(a,d) suggests that the Graphene interfacing effect contributes more to the overall magnetization than V-doping alone. The combined effect of V-doping and \ch{MoS2}/Gr interfacing results in the highest observed $M_S$ in the V-doped \ch{MoS2}/Gr HST (see the inset of Fig. \ref{fig:8}(c)), surpassing both the V-doped \ch{MoS2} monolayer and the \ch{MoS2}/Gr HST. These experimental observations are fully supported by our DFT simulations, which also predict that the presence of (C, S, Mo) vacancies in both Graphene and TMDC layers can enhance the overall magnetization, particularly when these defects are close and located near the interface. Electron transfer from \ch{MoS2} to Graphene further strengthens the FM coupling between V magnetic ions via p–d orbital hybridization (Fig. \ref{fig:7}(b)), contributing to the observed magnetization enhancement. Compared to the undoped HST, the charge transfer effect appears significantly stronger in the V-doped \ch{MoS2}/Gr system, especially in those with \ch{V_{Mo}} and \ch{V_{Mo+2S}} vacancies. Our combined theoretical and experimental findings not only validate the magnetization enhancement predicted for Mn-doped \ch{MoS2}/Gr heterostructures \cite{Tan2018}, but also provide new insights into the underlying physical mechanisms driving FM ordering in the 2D TMDC/Gr systems adding new functionalities to 2D van der Waals heterostructure-based nanodevices \cite{Safeer2019,Yu2014}.

\section{\label{sec:3} Conclusions}

The present study reports the geometric, electronic and magnetic properties of V-doped \ch{MoS2}/Gr HSTs with 2 and 4$\%$ V doping. All considered HSTs are metallic in nature with charge transfer from the Gr layer to the TMDC layer. This type of charge transfer is enhanced in the defective HSTs with a significant portion of the charge density localized near the defect sites.

We further find that Gr/\ch{MoS2} is nonmagnetic when doped with 2$\%$ V and becomes ferromagnetic with a total magnetic moment of 3 $\mu_B$/cell as the V doping concentration increases. The V-doped \ch{MoS2}/Gr HST with 4$\%$ V doping experiences stronger charge transfer as well as an upshift of the Dirac point. The effects of various defects, such as C monovacancy and Mo, S vacancies on the structural, electronic, and magnetic properties of V-doped \ch{MoS2}/Gr HSTs with 2$\%$ V doping are also studied. The overall geometry and stability of HSTs change insignificantly in the presence of defects. However, the defects in both Gr and V-doped \ch{MoS2} control the magnetic properties of the resulting HSTs. Pristine Gr remains nonmagnetic in all studied HSTs, while defective Gr displays strong magnetic properties. On the other hand, S mono and divacancies have secondary effects on the overall magnetization in the Gr/TMDC and (Gr+\ch{V_C})/TMDC HSTs, respectively. In contrast, the Mo monovacancy and (Mo+2S) complex defect in the $2\%$ doped TMDC noticeably affect the overall magnetization in the Gr/TMDC and (Gr+\ch{V_C})/TMDC HSTs, respectively. 

This study sheds light on the complex magnetism in sharp interfaced HSTs composed of Graphene and a TMDC mononlayer when doping and defects are taken into account. Tuning ferromagnetism in 2D systems by varying doping concentration and defect engineering is an attractive pathway for targeted applications in spintronics and other areas.

\section{\label{sec:4} Methods}

\subsection{Computational methods}
The first-principles calculations are performed in the Vienna ab-initio simulation package (VASP) \cite{Kresse1996} based on density functional theory (DFT) \cite{Hohenberg1964} and the projector augmented wave (PAW) method \cite{Kresse1999}. The PAW potentials describe the core states of C, Mo, S and V by the electronic configurations of [He]2s$^2$2p$^2$, [Ar]4s$^2$3d$^{10}$, [Ne]3s$^2$3p$^4$ and [Ar]3d$^3$4s$^2$, respectively. The Perdew-Burke-Ernzerhof (PBE) parameterization of the generalized gradient approximation (GGA) is employed to describe the exchange–correlation function \cite{Perdew1996}. The structures have been optimized using the conjugate gradient method with the forces calculated using the Hellman-Feynman theorem. The convergence criteria for the total energy and residual force on each atom are set to $10^{-5}$ eV and $10^{-2}$ eV/\AA, respectively. To isolate the interaction between the HST and its periodic images, a vacuum layer with a thickness of 15 \AA \text{} is inserted along the direction perpendicular to the layers. In addition, the DFT-D3(BJ) approach \cite{Grimme2011} is used to correct the vdW interaction in the HSTs. Considering the strong correlation effect of the V-3d electrons, a simplified DFT+U approach \cite{Dudarev1998} is employed with $U_{eff} = 3$ eV \cite{Thi2022}. The plane-wave energy cutoff is set to 500 eV, and a $\Gamma$-centered k meshes of $2 \times 2 \times 1$ is adopted to sample the first Brillouin zone for geometry optimization. The total energies and electronic structures are calculated with the optimized structures on a $9 \times 9 \times 1$ Monkhorst-Pack k-grid. The planar-averaged differential charge density offers a quantitative base for analyzing interfacial charge redistribution, which is defined as $\Delta \rho = \rho_{L1/L2} - \rho_{L1} - \rho_{L2}$ where $\rho_{L1/L2}$, $\rho_{L1}$ and $\rho_{L2}$ are the planar-averaged charge density of the HST and the isolated monolayers constituted the HST, respectively. 

\subsection{Experimental methods}
Pristine \ch{MoS2} and V-doped \ch{MoS2} monolayers were synthesized via chemical vapor deposition (CVD) and subsequently wet-transferred onto a CVD-grown graphene monolayer supported on a \ch{SiO2}/Si substrate. A Vanadium doping concentration of 2 at$\%$ was selected for this study, consistent with our previous work \cite{Zhou2024}. The Graphene monolayer was prepared through the thermal decomposition of a carbon-rich precursor, leading to the deposition of carbon atoms in a honeycomb lattice atop a copper film, which served as the catalytic substrate. After growth, the graphene layer was coated with polymethyl methacrylate (PMMA), and the underlying copper was etched away prior to transfer onto the \ch{SiO2}/Si substrate. The \ch{MoS2}/Gr and V-doped \ch{MoS2}/Gr HSTs were annealed at 300$^0$C for 10 hours to remove PMMA residues introduced during the transfer process. Post-annealing, the interlayer spacing between the V-doped \ch{MoS2} and Graphene was measured to be approximately 1.5 nm. For comparison, pristine and V-doped V-doped \ch{MoS2} monolayers were also annealed under the same conditions. 

The structural properties of both pristine and V-doped TMDC monolayers, as well as their corresponding TMDC/Graphene HSTs, were characterized using scanning transmission electron microscopy (STEM) and atomic force microscopy (AFM). The optical responses were examined via Raman and photoluminescence (PL) spectroscopy. Magnetic properties were measured using a vibrating sample magnetometry (VSM) probe integrated into a physical property measurement system (PPMS), under magnetic fields up to 9 T and across a broad temperature range from 2 K to 400 K.

\begin{acknowledgement}
D.T-X.D. acknowledges support from Presidential Fellowship sponsored by University of South Florida. L.M.W. acknowledges financial support from the US Department of Energy under Grant No.DE-FG02-06ER46297. Computational resources are provided by USF Research Computing. M.H.P acknowledgs support from the US Department of Energy, Office of Basic Energy Sciences, Division of Materials Science and Engineering under Grant No. DE-FG02-07ER46438 (magnetic measurements). 
\end{acknowledgement}

\section*{\label{sec:6}Author information}
\subsection*{Corresponding Author}
\begin{itemize}

    \item Manh-Huong Phan - Department of Physics, University of South Florida, Tampa, Florida 33620, United States; \href{https://orcid.org/0000-0002-6270-8990}{orcid.org/0000-0002-6270-8990} Email: \href{mailto:phanm@usf.edu}{phanm@usf.edu}.

    \item Lilia M. Woods - Department of Physics, University of South Florida, Tampa, Florida 33620, United States; \href{https://orcid.org/0000-0002-9872-1847}{orcid.org/0000-0002-9872-1847}; Email: \href{mailto:lmwoods@usf.edu}{lmwoods@usf.edu}.
\end{itemize}

\subsection*{Authors}

\begin{itemize}
    \item Diem Thi-Xuan Dang - Department of Physics, University of South Florida, Tampa, Florida 33620, United States; \href{https://orcid.org/0000-0001-7136-4125}{orcid.org/0000-0001-7136-4125}.

    \item Yen Thi-Hai Pham - Department of Physics, University of South Florida, Tampa, Florida 33620, United States; Current address: Department of Physics and Astronomy, George Mason University, Fairfax, Virginia 22030, USA.

    \item Da Zhou - Department of Physics \& Center for 2-Dimensional and Layered Materials, The Pennsylvania State University, University Park, Pennsylvania 16802, USA.

    \item Dai-Nam Le - Department of Physics, University of South Florida, Tampa, Florida 33620, United States; \href{https://orcid.org/0000-0003-0756-8742}{orcid.org/0000-0003-0756-8742}.

    \item Mauricio Terrones - Department of Physics \& Center for 2- Dimensional and Layered Materials, The Pennsylvania State University, University Park, Pennsylvania 16802, USA.

\end{itemize}

\section*{Conflicts of Interest}
There are no conflicts to declare.

\section*{Data Availability}
All data that support the findings of this study have been included in this publication. 

\bibliography{ref}

\providecommand{\latin}[1]{#1}
\makeatletter
\providecommand{\doi}
  {\begingroup\let\do\@makeother\dospecials
  \catcode`\{=1 \catcode`\}=2 \doi@aux}
\providecommand{\doi@aux}[1]{\endgroup\texttt{#1}}
\makeatother
\providecommand*\mcitethebibliography{\thebibliography}
\csname @ifundefined\endcsname{endmcitethebibliography}
  {\let\endmcitethebibliography\endthebibliography}{}
\begin{mcitethebibliography}{42}
\providecommand*\natexlab[1]{#1}
\providecommand*\mciteSetBstSublistMode[1]{}
\providecommand*\mciteSetBstMaxWidthForm[2]{}
\providecommand*\mciteBstWouldAddEndPuncttrue
  {\def\EndOfBibitem{\unskip.}}
\providecommand*\mciteBstWouldAddEndPunctfalse
  {\let\EndOfBibitem\relax}
\providecommand*\mciteSetBstMidEndSepPunct[3]{}
\providecommand*\mciteSetBstSublistLabelBeginEnd[3]{}
\providecommand*\EndOfBibitem{}
\mciteSetBstSublistMode{f}
\mciteSetBstMaxWidthForm{subitem}{(\alph{mcitesubitemcount})}
\mciteSetBstSublistLabelBeginEnd
  {\mcitemaxwidthsubitemform\space}
  {\relax}
  {\relax}

\bibitem[Luo \latin{et~al.}(2017)Luo, Xu, Zhu, Wu, McCormick, Zhan, Neupane,
  and Kawakami]{Luo2017}
Luo,~Y.~K.; Xu,~J.; Zhu,~T.; Wu,~G.; McCormick,~E.~J.; Zhan,~W.;
  Neupane,~M.~R.; Kawakami,~R.~K. Opto-valleytronic spin injection in monolayer
  \ch{MoS2}/few-layer graphene hybrid spin valves. \emph{Nano Letters}
  \textbf{2017}, \emph{17}, 3877--3883, DOI:
  \doi{10.1021/acs.nanolett.7b01393}\relax
\mciteBstWouldAddEndPuncttrue
\mciteSetBstMidEndSepPunct{\mcitedefaultmidpunct}
{\mcitedefaultendpunct}{\mcitedefaultseppunct}\relax
\EndOfBibitem
\bibitem[Avsar \latin{et~al.}(2017)Avsar, Unuchek, Liu, Sanchez, Watanabe,
  Taniguchi, Ozyilmaz, and Kis]{Avsar2017}
Avsar,~A.; Unuchek,~D.; Liu,~J.; Sanchez,~O.~L.; Watanabe,~K.; Taniguchi,~T.;
  Ozyilmaz,~B.; Kis,~A. Optospintronics in graphene via proximity coupling.
  \emph{ACS Nano} \textbf{2017}, \emph{11}, 11678--11686, DOI:
  \doi{10.1021/acsnano.7b06800}\relax
\mciteBstWouldAddEndPuncttrue
\mciteSetBstMidEndSepPunct{\mcitedefaultmidpunct}
{\mcitedefaultendpunct}{\mcitedefaultseppunct}\relax
\EndOfBibitem
\bibitem[Ghiasi \latin{et~al.}(2019)Ghiasi, Kaverzin, Blah, and
  Van~Wees]{Ghiasi2019}
Ghiasi,~T.~S.; Kaverzin,~A.~A.; Blah,~P.~J.; Van~Wees,~B.~J. Charge-to-spin
  conversion by the Rashba--Edelstein effect in two-dimensional van der Waals
  heterostructures up to room temperature. \emph{Nano Letters} \textbf{2019},
  \emph{19}, 5959--5966, DOI: \doi{10.1021/acs.nanolett.9b01611}\relax
\mciteBstWouldAddEndPuncttrue
\mciteSetBstMidEndSepPunct{\mcitedefaultmidpunct}
{\mcitedefaultendpunct}{\mcitedefaultseppunct}\relax
\EndOfBibitem
\bibitem[Benitez \latin{et~al.}(2020)Benitez, Savero~Torres, Sierra,
  Timmermans, Garcia, Roche, Costache, and Valenzuela]{Benitez2020}
Benitez,~L.~A.; Savero~Torres,~W.; Sierra,~J.~F.; Timmermans,~M.;
  Garcia,~J.~H.; Roche,~S.; Costache,~M.~V.; Valenzuela,~S.~O. Tunable
  room-temperature spin galvanic and spin Hall effects in van der Waals
  heterostructures. \emph{Nature Materials} \textbf{2020}, \emph{19}, 170--175,
  DOI: \doi{10.1038/s41563-019-0575-1}\relax
\mciteBstWouldAddEndPuncttrue
\mciteSetBstMidEndSepPunct{\mcitedefaultmidpunct}
{\mcitedefaultendpunct}{\mcitedefaultseppunct}\relax
\EndOfBibitem
\bibitem[Zhang \latin{et~al.}(2020)Zhang, Zheng, Sebastian, Olson, Liu,
  Fujisawa, Pham, Jimenez, Kalappattil, Miao, \latin{et~al.} others]{Zhang2020}
Zhang,~F.; Zheng,~B.; Sebastian,~A.; Olson,~D.~H.; Liu,~M.; Fujisawa,~K.;
  Pham,~Y. T.~H.; Jimenez,~V.~O.; Kalappattil,~V.; Miao,~L.; others Monolayer
  vanadium-doped tungsten disulfide: a room-temperature dilute magnetic
  semiconductor. \emph{Advanced Science} \textbf{2020}, \emph{7}, 2001174, DOI:
  \doi{10.1002/advs.202001174}\relax
\mciteBstWouldAddEndPuncttrue
\mciteSetBstMidEndSepPunct{\mcitedefaultmidpunct}
{\mcitedefaultendpunct}{\mcitedefaultseppunct}\relax
\EndOfBibitem
\bibitem[Coelho(2024)]{Coelho2024}
Coelho,~P.~M. Magnetic doping in transition metal dichalcogenides.
  \emph{Journal of Physics: Condensed Matter} \textbf{2024}, \emph{36}, 203001,
  DOI: \doi{10.1088/1361-648X/ad271b}\relax
\mciteBstWouldAddEndPuncttrue
\mciteSetBstMidEndSepPunct{\mcitedefaultmidpunct}
{\mcitedefaultendpunct}{\mcitedefaultseppunct}\relax
\EndOfBibitem
\bibitem[Pham \latin{et~al.}(2020)Pham, Liu, Jimenez, Yu, Kalappattil, Zhang,
  Wang, Williams, Terrones, and Phan]{Pham2020}
Pham,~Y. T.~H.; Liu,~M.; Jimenez,~V.~O.; Yu,~Z.; Kalappattil,~V.; Zhang,~F.;
  Wang,~K.; Williams,~T.; Terrones,~M.; Phan,~M.-H. Tunable ferromagnetism and
  thermally induced spin flip in Vanadium-doped tungsten diselenide monolayers
  at room temperature. \emph{Advanced Materials} \textbf{2020}, \emph{32},
  2003607, DOI: \doi{10.1002/adma.202003607}\relax
\mciteBstWouldAddEndPuncttrue
\mciteSetBstMidEndSepPunct{\mcitedefaultmidpunct}
{\mcitedefaultendpunct}{\mcitedefaultseppunct}\relax
\EndOfBibitem
\bibitem[Ortiz~Jimenez \latin{et~al.}(2021)Ortiz~Jimenez, Pham, Liu, Zhang, Yu,
  Kalappattil, Muchharla, Eggers, Duong, Terrones, \latin{et~al.}
  others]{Ortiz2021}
Ortiz~Jimenez,~V.; Pham,~Y. T.~H.; Liu,~M.; Zhang,~F.; Yu,~Z.; Kalappattil,~V.;
  Muchharla,~B.; Eggers,~T.; Duong,~D.~L.; Terrones,~M.; others
  Light-controlled room temperature ferromagnetism in vanadium-doped tungsten
  disulfide semiconducting monolayers. \emph{Advanced Electronic Materials}
  \textbf{2021}, \emph{7}, 2100030, DOI: \doi{10.1002/aelm.202100030}\relax
\mciteBstWouldAddEndPuncttrue
\mciteSetBstMidEndSepPunct{\mcitedefaultmidpunct}
{\mcitedefaultendpunct}{\mcitedefaultseppunct}\relax
\EndOfBibitem
\bibitem[Yun \latin{et~al.}(2020)Yun, Duong, Ha, Singh, Phan, Choi, Kim, and
  Lee]{Yun2020}
Yun,~S.~J.; Duong,~D.~L.; Ha,~D.~M.; Singh,~K.; Phan,~T.~L.; Choi,~W.;
  Kim,~Y.-M.; Lee,~Y.~H. Ferromagnetic Order at Room Temperature in Monolayer
  WSe2 Semiconductor via Vanadium Dopant. \emph{Advanced Science}
  \textbf{2020}, \emph{7}, 1903076, DOI:
  \doi{https://doi.org/10.1002/advs.201903076}\relax
\mciteBstWouldAddEndPuncttrue
\mciteSetBstMidEndSepPunct{\mcitedefaultmidpunct}
{\mcitedefaultendpunct}{\mcitedefaultseppunct}\relax
\EndOfBibitem
\bibitem[Nguyen \latin{et~al.}(2023)Nguyen, Jiang, Nguyen, Kim, Joo, Duong, and
  Lee]{Nguyen2023}
Nguyen,~L.-A.~T.; Jiang,~J.; Nguyen,~T.~D.; Kim,~P.; Joo,~M.-K.; Duong,~D.~L.;
  Lee,~Y.~H. Electrically tunable magnetic fluctuations in multilayered
  vanadium-doped tungsten diselenide. \emph{Nature Electronics} \textbf{2023},
  \emph{6}, 582--589, DOI:
  \doi{https://doi.org/10.1038/s41928-023-01002-1}\relax
\mciteBstWouldAddEndPuncttrue
\mciteSetBstMidEndSepPunct{\mcitedefaultmidpunct}
{\mcitedefaultendpunct}{\mcitedefaultseppunct}\relax
\EndOfBibitem
\bibitem[Zhang \latin{et~al.}(2020)Zhang, Zhu, Tebyetekerwa, Li, Liu, Lei,
  Wang, Zhang, and Lu]{Zhang2020_2}
Zhang,~J.; Zhu,~Y.; Tebyetekerwa,~M.; Li,~D.; Liu,~D.; Lei,~W.; Wang,~L.;
  Zhang,~Y.; Lu,~Y. Vanadium-doped monolayer \ch{MoS2} with tunable optical
  properties for field-effect transistors. \emph{ACS Applied Nano Materials}
  \textbf{2020}, \emph{4}, 769--777, DOI: \doi{10.1021/acsanm.0c03083}\relax
\mciteBstWouldAddEndPuncttrue
\mciteSetBstMidEndSepPunct{\mcitedefaultmidpunct}
{\mcitedefaultendpunct}{\mcitedefaultseppunct}\relax
\EndOfBibitem
\bibitem[Sahoo \latin{et~al.}(2022)Sahoo, Panda, Bawari, Sharma, Maity, Lal,
  Arenal, Rajalaksmi, and Narayanan]{Sahoo2022}
Sahoo,~K.~R.; Panda,~J.~J.; Bawari,~S.; Sharma,~R.; Maity,~D.; Lal,~A.;
  Arenal,~R.; Rajalaksmi,~G.; Narayanan,~T.~N. Enhanced room-temperature
  spin-valley coupling in V-doped \ch{MoS2}. \emph{Physical Review Materials}
  \textbf{2022}, \emph{6}, 085202, DOI:
  \doi{10.1103/PhysRevMaterials.6.085202}\relax
\mciteBstWouldAddEndPuncttrue
\mciteSetBstMidEndSepPunct{\mcitedefaultmidpunct}
{\mcitedefaultendpunct}{\mcitedefaultseppunct}\relax
\EndOfBibitem
\bibitem[Maity \latin{et~al.}(2022)Maity, Sharma, Sahoo, Lal, Arenal, and
  Narayanan]{Maity2022}
Maity,~D.; Sharma,~R.; Sahoo,~K.~R.; Lal,~A.; Arenal,~R.; Narayanan,~T.~N. On
  the existence of photoluminescence and room-temperature spin polarization in
  ambipolar V doped \ch{MoS2} monolayers. \emph{arXiv preprint
  arXiv:2204.05887} \textbf{2022}, \relax
\mciteBstWouldAddEndPunctfalse
\mciteSetBstMidEndSepPunct{\mcitedefaultmidpunct}
{}{\mcitedefaultseppunct}\relax
\EndOfBibitem
\bibitem[Zhou \latin{et~al.}(2024)Zhou, Pham, Dang, Sanchez, Oberoi, Wang,
  Fest, Sredenschek, Liu, Terrones, Das, Le, Woods, Phan, and
  Terrones]{Zhou2024}
Zhou,~D.; Pham,~Y. T.~H.; Dang,~D. T.-X.; Sanchez,~D.; Oberoi,~A.; Wang,~K.;
  Fest,~A.; Sredenschek,~A.; Liu,~M.; Terrones,~H.; Das,~S.; Le,~D.-N.;
  Woods,~L.~M.; Phan,~M.-H.; Terrones,~M. Vanadium-doped molybdenum disulfide
  monolayers with tunable electronic and magnetic properties: do
  vanadium-vacancy pairs matter? \emph{arXiv preprint arXiv:2401.16806}
  \textbf{2024}, \relax
\mciteBstWouldAddEndPunctfalse
\mciteSetBstMidEndSepPunct{\mcitedefaultmidpunct}
{}{\mcitedefaultseppunct}\relax
\EndOfBibitem
\bibitem[Tan \latin{et~al.}(2018)Tan, Wang, Liu, Liu, Feng, and Yu]{Tan2018}
Tan,~Q.; Wang,~Q.; Liu,~Y.; Liu,~C.; Feng,~X.; Yu,~D. Enhanced magnetic
  properties and tunable Dirac point of graphene/Mn-doped monolayer MoS2
  heterostructures. \emph{Journal of Physics: Condensed Matter} \textbf{2018},
  \emph{30}, 305304, DOI: \doi{10.1088/1361-648X/aacca2}\relax
\mciteBstWouldAddEndPuncttrue
\mciteSetBstMidEndSepPunct{\mcitedefaultmidpunct}
{\mcitedefaultendpunct}{\mcitedefaultseppunct}\relax
\EndOfBibitem
\bibitem[Cai \latin{et~al.}(2021)Cai, Duan, Liu, Wang, Tan, Hu, Hu, Sun, and
  Yan]{Cai2021}
Cai,~L.; Duan,~H.; Liu,~Q.; Wang,~C.; Tan,~H.; Hu,~W.; Hu,~F.; Sun,~Z.; Yan,~W.
  Ultrahigh-temperature ferromagnetism in MoS2 Moiré superlattice/graphene
  hybrid heterostructures. \emph{Nano Research} \textbf{2021}, \emph{14},
  4182--4187, DOI: \doi{https://doi.org/10.1007/s12274-021-3360-9}\relax
\mciteBstWouldAddEndPuncttrue
\mciteSetBstMidEndSepPunct{\mcitedefaultmidpunct}
{\mcitedefaultendpunct}{\mcitedefaultseppunct}\relax
\EndOfBibitem
\bibitem[Nakhmedov \latin{et~al.}(2019)Nakhmedov, Nadimi, Vedaei, Alekperov,
  Tatardar, Najafov, Abbasov, and Saletsky]{Nakhmedov2019}
Nakhmedov,~E.; Nadimi,~E.; Vedaei,~S.; Alekperov,~O.; Tatardar,~F.;
  Najafov,~A.; Abbasov,~I.; Saletsky,~A. Vacancy mediated magnetization and
  healing of a graphene monolayer. \emph{Physical Review B} \textbf{2019},
  \emph{99}, 125125, DOI: \doi{10.1103/PhysRevB.99.125125}\relax
\mciteBstWouldAddEndPuncttrue
\mciteSetBstMidEndSepPunct{\mcitedefaultmidpunct}
{\mcitedefaultendpunct}{\mcitedefaultseppunct}\relax
\EndOfBibitem
\bibitem[Tiwari \latin{et~al.}(2023)Tiwari, Pandey, Pandey, Wang,
  Bystrzejewski, Mishra, and Zhu]{Tiwari2023}
Tiwari,~S.~K.; Pandey,~S.~K.; Pandey,~R.; Wang,~N.; Bystrzejewski,~M.;
  Mishra,~Y.~K.; Zhu,~Y. Stone--wales defect in graphene. \emph{Small}
  \textbf{2023}, \emph{19}, 2303340, DOI: \doi{10.1002%2Fsmll.202303340}\relax
\mciteBstWouldAddEndPuncttrue
\mciteSetBstMidEndSepPunct{\mcitedefaultmidpunct}
{\mcitedefaultendpunct}{\mcitedefaultseppunct}\relax
\EndOfBibitem
\bibitem[Lin \latin{et~al.}(2016)Lin, Carvalho, Kahn, Lv, Rao, Terrones,
  Pimenta, and Terrones]{Lin2016}
Lin,~Z.; Carvalho,~B.~R.; Kahn,~E.; Lv,~R.; Rao,~R.; Terrones,~H.;
  Pimenta,~M.~A.; Terrones,~M. Defect engineering of two-dimensional transition
  metal dichalcogenides. \emph{2D Materials} \textbf{2016}, \emph{3}, 022002,
  DOI: \doi{10.1088/2053-1583/3/2/022002}\relax
\mciteBstWouldAddEndPuncttrue
\mciteSetBstMidEndSepPunct{\mcitedefaultmidpunct}
{\mcitedefaultendpunct}{\mcitedefaultseppunct}\relax
\EndOfBibitem
\bibitem[Han \latin{et~al.}(2024)Han, Niu, Luo, Li, Dan, Hong, Wu, Trukhanov,
  Ji, Wang, \latin{et~al.} others]{Han2024}
Han,~X.; Niu,~M.; Luo,~Y.; Li,~R.; Dan,~J.; Hong,~Y.; Wu,~X.; Trukhanov,~A.~V.;
  Ji,~W.; Wang,~Y.; others Atomically engineering metal vacancies in monolayer
  transition metal dichalcogenides. \emph{Nature Synthesis} \textbf{2024},
  \emph{3}, 586--594, DOI: \doi{10.1038/s44160-024-00501-z}\relax
\mciteBstWouldAddEndPuncttrue
\mciteSetBstMidEndSepPunct{\mcitedefaultmidpunct}
{\mcitedefaultendpunct}{\mcitedefaultseppunct}\relax
\EndOfBibitem
\bibitem[Khan \latin{et~al.}(2017)Khan, Erementchouk, Hendrickson, and
  Leuenberger]{Khan2017}
Khan,~M.; Erementchouk,~M.; Hendrickson,~J.; Leuenberger,~M.~N. Electronic and
  optical properties of vacancy defects in single-layer transition metal
  dichalcogenides. \emph{Physical Review B} \textbf{2017}, \emph{95}, 245435,
  DOI: \doi{10.1103/PhysRevB.95.245435}\relax
\mciteBstWouldAddEndPuncttrue
\mciteSetBstMidEndSepPunct{\mcitedefaultmidpunct}
{\mcitedefaultendpunct}{\mcitedefaultseppunct}\relax
\EndOfBibitem
\bibitem[Ko{\'o}s \latin{et~al.}(2019)Ko{\'o}s, Vancs{\'o}, Szendrő, Dobrik,
  Antognini~Silva, Popov, Sorokin, Henrard, Hwang, Bir{\'o}, \latin{et~al.}
  others]{Koos2019}
Ko{\'o}s,~A.~A.; Vancs{\'o},~P.; Szendrő,~M.; Dobrik,~G.; Antognini~Silva,~D.;
  Popov,~Z.~I.; Sorokin,~P.~B.; Henrard,~L.; Hwang,~C.; Bir{\'o},~L.~P.; others
  Influence of native defects on the electronic and magnetic properties of CVD
  grown \ch{MoSe2} single layers. \emph{The Journal of Physical Chemistry C}
  \textbf{2019}, \emph{123}, 24855--24864, DOI:
  \doi{10.1021/acs.jpcc.9b05921}\relax
\mciteBstWouldAddEndPuncttrue
\mciteSetBstMidEndSepPunct{\mcitedefaultmidpunct}
{\mcitedefaultendpunct}{\mcitedefaultseppunct}\relax
\EndOfBibitem
\bibitem[Dang \latin{et~al.}(2025)Dang, Le, and Woods]{Dang2025}
Dang,~D. T.-X.; Le,~D.-N.; Woods,~L.~M. Dissecting van der Waals interactions
  with Density Functional Theory--Wannier-basis approach. \emph{Computer
  Physics Communications} \textbf{2025}, 109525, DOI:
  \doi{10.1016/j.cpc.2025.109525}\relax
\mciteBstWouldAddEndPuncttrue
\mciteSetBstMidEndSepPunct{\mcitedefaultmidpunct}
{\mcitedefaultendpunct}{\mcitedefaultseppunct}\relax
\EndOfBibitem
\bibitem[Wang \latin{et~al.}(2016)Wang, Li, and Yi]{Wang2016}
Wang,~Y.; Li,~S.; Yi,~J. Electronic and magnetic properties of Co doped
  \ch{MoS2} monolayer. \emph{Scientific Reports} \textbf{2016}, \emph{6},
  24153, DOI: \doi{10.1038/srep24153}\relax
\mciteBstWouldAddEndPuncttrue
\mciteSetBstMidEndSepPunct{\mcitedefaultmidpunct}
{\mcitedefaultendpunct}{\mcitedefaultseppunct}\relax
\EndOfBibitem
\bibitem[Hu \latin{et~al.}(2019)Hu, Tan, Duan, Li, Li, Ji, Lu, Wang, Sun, Hu,
  Wang, and Yan]{Hu2019}
Hu,~W.; Tan,~H.; Duan,~H.; Li,~G.; Li,~N.; Ji,~Q.; Lu,~Y.; Wang,~Y.; Sun,~Z.;
  Hu,~F.; Wang,~C.; Yan,~W. Synergetic Effect of Substitutional Dopants and
  Sulfur Vacancy in Modulating the Ferromagnetism of MoS2 Nanosheets. \emph{ACS
  Applied Materials \& Interfaces} \textbf{2019}, \emph{11}, 31155–31161,
  DOI: \doi{https://doi.org/10.1021/acsami.9b09165}\relax
\mciteBstWouldAddEndPuncttrue
\mciteSetBstMidEndSepPunct{\mcitedefaultmidpunct}
{\mcitedefaultendpunct}{\mcitedefaultseppunct}\relax
\EndOfBibitem
\bibitem[Lieb(1989)]{Lieb1989}
Lieb,~E.~H. Two theorems on the Hubbard model. \emph{Physical Review Letters}
  \textbf{1989}, \emph{62}, 1201, DOI: \doi{10.1103/PhysRevLett.62.1201}\relax
\mciteBstWouldAddEndPuncttrue
\mciteSetBstMidEndSepPunct{\mcitedefaultmidpunct}
{\mcitedefaultendpunct}{\mcitedefaultseppunct}\relax
\EndOfBibitem
\bibitem[Yazyev and Helm(2007)Yazyev, and Helm]{Yazyev2007}
Yazyev,~O.~V.; Helm,~L. Defect-induced magnetism in graphene. \emph{Physical
  Review B—Condensed Matter and Materials Physics} \textbf{2007}, \emph{75},
  125408, DOI: \doi{10.1103/PhysRevB.75.125408}\relax
\mciteBstWouldAddEndPuncttrue
\mciteSetBstMidEndSepPunct{\mcitedefaultmidpunct}
{\mcitedefaultendpunct}{\mcitedefaultseppunct}\relax
\EndOfBibitem
\bibitem[Thi-Xuan~Dang \latin{et~al.}(2022)Thi-Xuan~Dang, Barik, Phan, and
  Woods]{Thi2022}
Thi-Xuan~Dang,~D.; Barik,~R.~K.; Phan,~M.-H.; Woods,~L.~M. Enhanced magnetism
  in heterostructures with transition-metal dichalcogenide monolayers.
  \emph{The Journal of Physical Chemistry Letters} \textbf{2022}, \emph{13},
  8879--8887, DOI: \doi{10.1021/acs.jpclett.2c01925}\relax
\mciteBstWouldAddEndPuncttrue
\mciteSetBstMidEndSepPunct{\mcitedefaultmidpunct}
{\mcitedefaultendpunct}{\mcitedefaultseppunct}\relax
\EndOfBibitem
\bibitem[Hung \latin{et~al.}(2023)Hung, Dang, Chanda, Detellem, Alzahrani,
  Kapuruge, Pham, Liu, Zhou, Gutierrez, Arena, Terrones, Witanachchi, Woods,
  Srikanth, and Phan]{Hung2023}
Hung,~C.-M. \latin{et~al.}  Enhanced magnetism and anomalous Hall transport
  through two-dimensional tungsten disulfide interfaces. \emph{Nanomaterials}
  \textbf{2023}, \emph{13}, DOI: \doi{10.3390/nano13040771}\relax
\mciteBstWouldAddEndPuncttrue
\mciteSetBstMidEndSepPunct{\mcitedefaultmidpunct}
{\mcitedefaultendpunct}{\mcitedefaultseppunct}\relax
\EndOfBibitem
\bibitem[Pierucci \latin{et~al.}(2016)Pierucci, Henck, Avila, Balan, Naylor,
  Patriarche, Dappe, Silly, Sirotti, Johnson, Asensio, and
  Ouerghi]{Pierucci2016}
Pierucci,~D.; Henck,~H.; Avila,~J.; Balan,~A.; Naylor,~C.~H.; Patriarche,~G.;
  Dappe,~Y.~J.; Silly,~M.~G.; Sirotti,~F.; Johnson,~A. T.~C.; Asensio,~M.~C.;
  Ouerghi,~A. Band alignment and minigaps in monolayer \ch{MoS2}-Graphene van
  der Waals heterostructures. \emph{Nano Letters} \textbf{2016}, \emph{16},
  4054--4061, DOI: \doi{10.1021/acs.nanolett.6b00609}\relax
\mciteBstWouldAddEndPuncttrue
\mciteSetBstMidEndSepPunct{\mcitedefaultmidpunct}
{\mcitedefaultendpunct}{\mcitedefaultseppunct}\relax
\EndOfBibitem
\bibitem[Chitara \latin{et~al.}(2023)Chitara, Dimitrov, Liu, Seling, Kolli,
  Zhou, Yu, Shringi, Terrones, and Yan]{Chitara2023}
Chitara,~B.; Dimitrov,~E.; Liu,~M.; Seling,~T.~R.; Kolli,~B.~S.; Zhou,~D.;
  Yu,~Z.; Shringi,~A.~K.; Terrones,~M.; Yan,~F. Charge transfer modulation in
  Vanadium-doped \ch{WS2}/\ch{Bi2O2Se} heterostructures. \emph{Small}
  \textbf{2023}, \emph{19}, 2302289, DOI: \doi{10.1002/smll.202302289}\relax
\mciteBstWouldAddEndPuncttrue
\mciteSetBstMidEndSepPunct{\mcitedefaultmidpunct}
{\mcitedefaultendpunct}{\mcitedefaultseppunct}\relax
\EndOfBibitem
\bibitem[Cai \latin{et~al.}(2015)Cai, He, Liu, Yao, Chen, Yan, Hu, Jiang, Zhao,
  Hu, Sun, and Wei]{Cai2015}
Cai,~L.; He,~J.; Liu,~Q.; Yao,~T.; Chen,~L.; Yan,~W.; Hu,~F.; Jiang,~Y.;
  Zhao,~Y.; Hu,~T.; Sun,~Z.; Wei,~S. Vacancy-Induced Ferromagnetism of MoS2
  Nanosheets. \emph{J. Am. Chem. Soc.} \textbf{2015}, \emph{137}, 2622–2627,
  DOI: \doi{https://doi.org/10.1021/ja5120908}\relax
\mciteBstWouldAddEndPuncttrue
\mciteSetBstMidEndSepPunct{\mcitedefaultmidpunct}
{\mcitedefaultendpunct}{\mcitedefaultseppunct}\relax
\EndOfBibitem
\bibitem[Anbalagan \latin{et~al.}(2023)Anbalagan, Hu, Chan, Gandhi, Gupta,
  Chaudhary, Chuang, Ramesh, Billo, Sabbah, Chiang, Tseng, Chueh, Wu, Tai,
  Chen, and Lee]{Anbalagan2023}
Anbalagan,~A.~k. \latin{et~al.}  Gamma-Ray Irradiation Induced Ultrahigh
  Room-Temperature Ferromagnetism in MoS2 Sputtered Few-Layered Thin Films.
  \emph{ACS Nano} \textbf{2023}, \emph{17}, 6555–6564, DOI:
  \doi{https://doi.org/10.1021/acsnano.2c11955}\relax
\mciteBstWouldAddEndPuncttrue
\mciteSetBstMidEndSepPunct{\mcitedefaultmidpunct}
{\mcitedefaultendpunct}{\mcitedefaultseppunct}\relax
\EndOfBibitem
\bibitem[Safeer \latin{et~al.}(2019)Safeer, Ingla-Aynés, Herling, Garcia,
  Vila, Ontoso, Calvo, Roche, Hueso, and Casanova]{Safeer2019}
Safeer,~C.~K.; Ingla-Aynés,~J.; Herling,~F.; Garcia,~J.~H.; Vila,~M.;
  Ontoso,~N.; Calvo,~M.~R.; Roche,~S.; Hueso,~L.~E.; Casanova,~F.
  Room-Temperature Spin Hall Effect in Graphene/MoS2 van der Waals
  Heterostructures. \emph{Nano Letters} \textbf{2019}, \emph{19}, 1074–1082,
  DOI: \doi{https://doi.org/10.1021/acs.nanolett.8b04368}\relax
\mciteBstWouldAddEndPuncttrue
\mciteSetBstMidEndSepPunct{\mcitedefaultmidpunct}
{\mcitedefaultendpunct}{\mcitedefaultseppunct}\relax
\EndOfBibitem
\bibitem[Yu \latin{et~al.}(2014)Yu, Lee, Ling, Santos, Shin, Lin, Dubey,
  Kaxiras, Kong, Wang, and Palacios]{Yu2014}
Yu,~L.; Lee,~Y.-H.; Ling,~X.; Santos,~E. J.~G.; Shin,~Y.~C.; Lin,~Y.;
  Dubey,~M.; Kaxiras,~E.; Kong,~J.; Wang,~H.; Palacios,~T. Graphene/MoS2 Hybrid
  Technology for Large-Scale Two-Dimensional Electronics. \emph{Nano Letters}
  \textbf{2014}, \emph{14}, 3055–3063, DOI:
  \doi{https://doi.org/10.1021/nl404795z}\relax
\mciteBstWouldAddEndPuncttrue
\mciteSetBstMidEndSepPunct{\mcitedefaultmidpunct}
{\mcitedefaultendpunct}{\mcitedefaultseppunct}\relax
\EndOfBibitem
\bibitem[Kresse and Furthm{\"u}ller(1996)Kresse, and
  Furthm{\"u}ller]{Kresse1996}
Kresse,~G.; Furthm{\"u}ller,~J. Efficient iterative schemes for ab initio
  total-energy calculations using a plane-wave basis set. \emph{Physical Review
  B} \textbf{1996}, \emph{54}, 11169, DOI:
  \doi{10.1103/PhysRevB.54.11169}\relax
\mciteBstWouldAddEndPuncttrue
\mciteSetBstMidEndSepPunct{\mcitedefaultmidpunct}
{\mcitedefaultendpunct}{\mcitedefaultseppunct}\relax
\EndOfBibitem
\bibitem[Hohenberg and Kohn(1964)Hohenberg, and Kohn]{Hohenberg1964}
Hohenberg,~P.; Kohn,~W. Inhomogeneous electron gas. \emph{Physical Review}
  \textbf{1964}, \emph{136}, B864, DOI: \doi{10.1103/PhysRev.136.B864}\relax
\mciteBstWouldAddEndPuncttrue
\mciteSetBstMidEndSepPunct{\mcitedefaultmidpunct}
{\mcitedefaultendpunct}{\mcitedefaultseppunct}\relax
\EndOfBibitem
\bibitem[Kresse and Joubert(1999)Kresse, and Joubert]{Kresse1999}
Kresse,~G.; Joubert,~D. From ultrasoft pseudopotentials to the projector
  augmented-wave method. \emph{Physical Review B} \textbf{1999}, \emph{59},
  1758, DOI: \doi{10.1103/PhysRevB.59.1758}\relax
\mciteBstWouldAddEndPuncttrue
\mciteSetBstMidEndSepPunct{\mcitedefaultmidpunct}
{\mcitedefaultendpunct}{\mcitedefaultseppunct}\relax
\EndOfBibitem
\bibitem[Perdew \latin{et~al.}(1996)Perdew, Burke, and Ernzerhof]{Perdew1996}
Perdew,~J.~P.; Burke,~K.; Ernzerhof,~M. Generalized gradient approximation made
  simple. \emph{Physical Review Letters} \textbf{1996}, \emph{77}, 3865, DOI:
  \doi{10.1103/PhysRevLett.77.3865}\relax
\mciteBstWouldAddEndPuncttrue
\mciteSetBstMidEndSepPunct{\mcitedefaultmidpunct}
{\mcitedefaultendpunct}{\mcitedefaultseppunct}\relax
\EndOfBibitem
\bibitem[Grimme \latin{et~al.}(2011)Grimme, Ehrlich, and Goerigk]{Grimme2011}
Grimme,~S.; Ehrlich,~S.; Goerigk,~L. Effect of the damping function in
  dispersion corrected density functional theory. \emph{Journal of
  Computational Chemistry} \textbf{2011}, \emph{32}, 1456--1465, DOI:
  \doi{10.1002/jcc.21759}\relax
\mciteBstWouldAddEndPuncttrue
\mciteSetBstMidEndSepPunct{\mcitedefaultmidpunct}
{\mcitedefaultendpunct}{\mcitedefaultseppunct}\relax
\EndOfBibitem
\bibitem[Dudarev \latin{et~al.}(1998)Dudarev, Botton, Savrasov, Humphreys, and
  Sutton]{Dudarev1998}
Dudarev,~S.~L.; Botton,~G.~A.; Savrasov,~S.~Y.; Humphreys,~C.; Sutton,~A.~P.
  Electron-energy-loss spectra and the structural stability of nickel oxide: An
  LSDA+ U study. \emph{Physical Review B} \textbf{1998}, \emph{57}, 1505, DOI:
  \doi{10.1103/PhysRevB.57.1505}\relax
\mciteBstWouldAddEndPuncttrue
\mciteSetBstMidEndSepPunct{\mcitedefaultmidpunct}
{\mcitedefaultendpunct}{\mcitedefaultseppunct}\relax
\EndOfBibitem
\end{mcitethebibliography}

\end{document}